\documentstyle[epsf]{jfm}

\input psfig.sty

\def\psimax{\psi_{\rmn max}}
\def\cosec{{\rmn cosec}}

\def\Q{{\cal Q}}
\def\R{{\cal R}}
\def\SS{{\cal S}}
\def\Ri{\mbox{\it Ri}}   
\def\Fr{\mbox{\it Fr}}   
\def\Real{\mbox{Re}}      
\def\Imag{\mbox{Im}}      
\def\Rey{\mbox{\it Re}}   
\def\Pen{\mbox{\it Pe}}   
\def\ii{{\rmn  i}} 

\catcode`\œ=\active \gdefœ{\setbox0=\hbox{0}\hbox to\wd0{}}
\def\Real{\mbox{Re}}      
\def\Imag{\mbox{Im}}      
\def\Rey{\mbox{\it Re}}   
\def\Pen{\mbox{\it Pe}}   
\ifCUPmtlplainloaded
\else
\fi
\ifCUPmtlplainloaded

\fi
\ifCUPmtlplainloaded
  \font\bit = mtmib10 at 10.5pt \skewchar\bit ='177  
\else
  \font\bit = cmmib10 \skewchar\bit ='177  
\fi
\ifCUPmtlplainloaded
\else
  \font\tenbmi=cmmib10 at 10pt  \skewchar\tenbmi ='177
  \font\sevenbmi=cmmib10 at 7pt \skewchar\sevenbmi ='177
  \font\fivebmi=cmmib10 at 5pt  \skewchar\fivebmi ='177

  \newfam\bmifam
  \textfont\bmifam=\tenbmi
  \scriptfont\bmifam=\sevenbmi
  \scriptscriptfont\bmifam=\fivebmi
  \def\bmi{\fam\bmifam\tenbmi}
\fi
\newsavebox{\thalfbox}
\sbox{\thalfbox}{$\textstyle\frac{1}{2}$}

\newsavebox{\shalfbox}
\sbox{\shalfbox}{$\scriptstyle\frac{1}{2}$}

\newsavebox{\squartbox}
\sbox{\squartbox}{$\frac{1}{4}$} 

\newsavebox{\etbox}
\sbox{\etbox}{\boldmath$\eta$}

\newsavebox{\astrutbox}
\sbox{\astrutbox}{\rule[-5pt]{0pt}{20pt}}

\ifnfsstwo
  
  \newcommand{\rmn}[1] {{\mathrm #1}}
  \newcommand{\itl}[1] {{\mathit #1}}
  
\fi
\ifnfssone
  \newmathalphabet{\mathit}
    \addtoversion{normal}{\mathit}{cmr}{m}{it}
    \addtoversion{bold}{\mathit}{cmr}{bx}{it}
  
  \newcommand{\rmn}[1] {{\mathrm #1}}
  \newcommand{\itl}[1] {{\mathit #1}}
  
\fi
\ifoldfss    
  
  \newcommand{\rmn}[1] {{\rm #1}}
  \newcommand{\itl}[1] {{\it #1}}
  
\fi

\mathchardef\varLambda="0103

\ifCUPmtlplainloaded
\else
\fi
\ifCUPmtlplainloaded
  \let\bcdot=\undefined
  \NewSymbolFont{bldsym}{mtbsy10}{'60}
  \NewMathSymbol{\bcdot}{2}{bldsym}{01}
\else
  \font\tenbms=cmbsy10          \skewchar\tenbms ='60
  \font\sevenbms=cmbsy10 at 7pt \skewchar\sevenbms ='60
  \font\fivebms=cmbsy10 at 5pt  \skewchar\fivebms ='60

  \newfam\bmsfam
  \textfont\bmsfam=\tenbms
  \scriptfont\bmsfam=\sevenbms
  \scriptscriptfont\bmsfam=\fivebms

  \edef\bsy{\hexnumber\bmsfam}
  \mathchardef\bnabla="0\bsy72
  \mathchardef\bcdotsymbol="0\bsy01
  \def\bcdot{\,\bcdotsymbol\,}
\fi
\def\etal{\mbox{\it et al.\ }}

\title[Near critical reflection of internal waves]
{Near critical reflection of internal waves}
\author[T. Dauxois and  W.R. Young]
{T\ls H\ls I\ls E\ls R\ls R\ls Y D\ls A\ls U\ls X\ls O\ls I\ls
S
\thanks{Address for correspondence: Laboratoire de Physique,
ENS Lyon, 
46 All\'{e}e d'Italie, 69007 Lyon, France 
(email: Thierry.Dauxois@ens-lyon.fr).}
\ns \and  W.\ns R.\ns Y\ls O\ls U\ls N\ls G
}
\affiliation{
Scripps Institution of Oceanography,
University of California at San Diego,\\
La Jolla, CA~92093--0230, USA}
\date{\today}

\begin{document}
\maketitle
\begin{abstract}

Using a matched asymptotic expansion we analyze the two-dimensional,
near-critical reflection of a weakly nonlinear, internal gravity wave
from a sloping boundary in a uniformly stratified fluid.  Taking a
distinguished limit in which the amplitude of the incident wave, the
dissipation, and the departure from criticality are all small, we
obtain a reduced description of the dynamics. This simplification
shows how either dissipation or transience heals the singularity which
is presented by the solution of Phillips
({\it The Dynamics of the Upper Ocean}, 1966) in the precisely
critical case. In the inviscid critical case, an explicit solution of
the initial value problem shows that the buoyancy perturbation and the
along-slope velocity both grow linearly with time, while the scale of
the reflected disturbance is reduced as $1/t$. During the course of
this scale reduction, the stratification is `overturned' and the
Miles-Howard condition for stratified shear flow stability is
violated. However, for all slope angles, the `overturning' occurs
before the Miles-Howard stability condition is violated and so we
argue that the first instability is convective.

Solutions of the simplified dynamics resemble 
 certain experimental visualizations of the reflection process. In
particular, the buoyancy field computed from the analytic solution is
in good agreement with  visualizations reported by
\cite{thorpehaines} {\it J. Fluid Mech.} {\bf 178}, 299-302.

One curious aspect of the weakly nonlinear theory is that the final reduced
description is a linear equation (at the solvability order in the expansion
all of the apparently resonant nonlinear contributions cancel amongst
themselves). However the reconstructed fields do contain nonlinearly driven
second harmonics  which are responsible for an important symmetry breaking
in which alternate vortices  differ in strength and
size from their immediate neighbours.

\end{abstract}

\section{Introduction}

\subsection{Reflection of internal waves on a sloping bottom}

Understanding  the intensity and spatial distribution of turbulent
vertical mixing in the ocean is an important problem in physical
oceanography.  Ocean models require
accurate parametrizations of turbulent mixing to make realistic
predictions of the transport of heat, salt and chemical species.
Because the ocean is stably stratified, vertical mixing is
inhibited and  convection to great depth occurs only in
restricted high latitude regions. But the fluid
which has reached the abyss by convection must ultimately return to the
sea surface so as to maintain a quasi-steady state.  How and where this
return flow occurs remains obscure to this day.
\cite{munk} and \cite{armi} suggested that significant vertical mixing
takes place at sloping boundaries. Indeed, a recent report by
\cite{polzin97} shows greatly elevated mixing rates  above rough
topography, especially in the deepest 150 m of the Brazil basin.

\begin{figure}\null\hskip3truecm
\psfig{figure=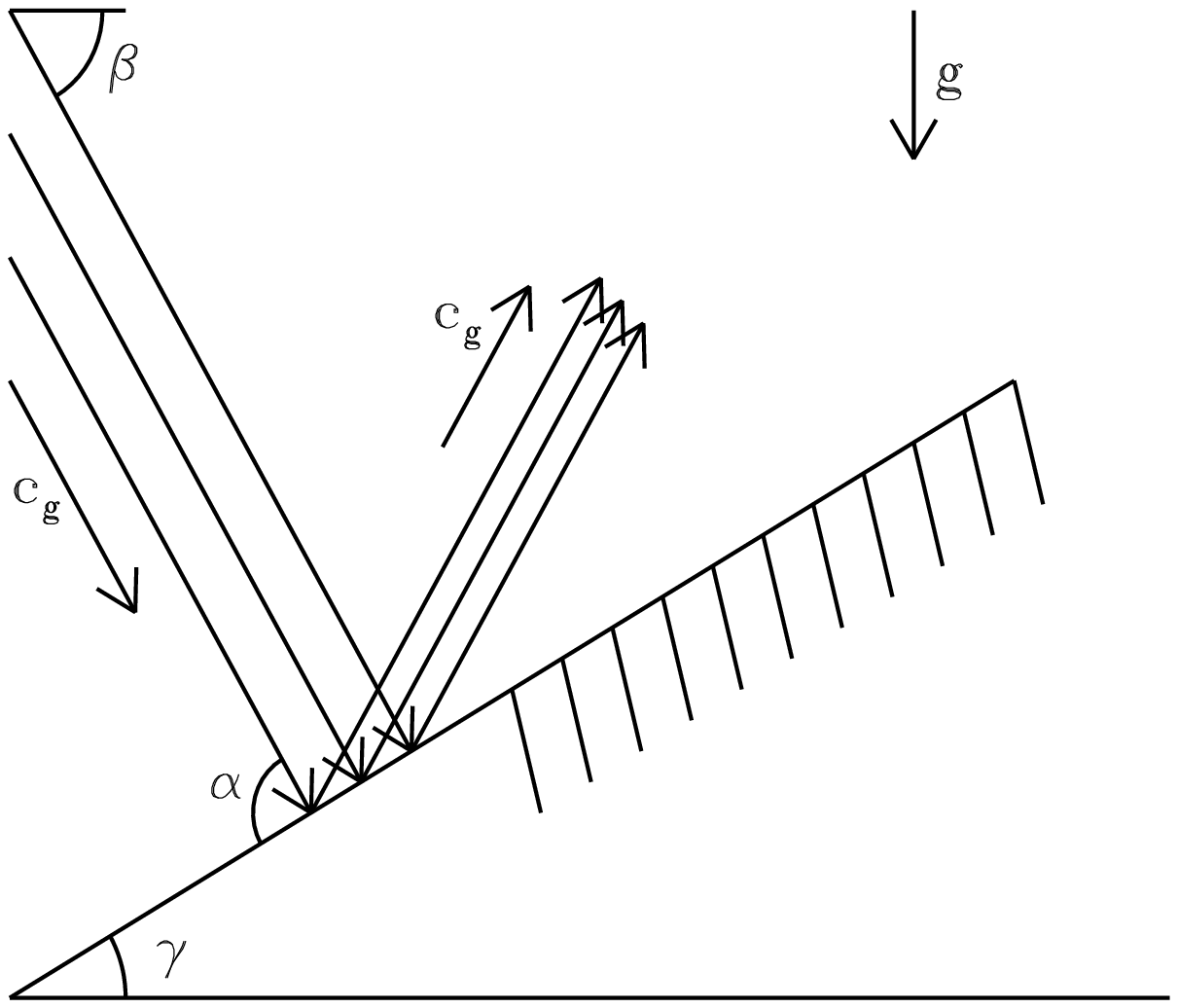,height=15truecm,width=15truecm}
\caption{Schematic view of the reflection of the incident wave. The angle
between the bottom slope and the horizontal is $\gamma$;
the  angle between the incident group velocity
and the horizontal is $\beta$, and $\alpha=\beta+\gamma$.
${\bmi c}_g$ indicates the group velocity and ${\bmi g}$ indicates gravity.}
\label{geometry}
\end{figure}

\cite{sandstrom66} was the first to propose  the oceanic internal wave
field as a possible source of the energy which is needed to activate
strong mixing near sloping boundaries. Internal waves are well
documented in nature (see Munk 1981) and they obey simple, but
unusual, reflection laws at a rigid boundary. In optics or acoustics,
the incident and reflected wave rays make the same angle with respect to
the {\em normal} to the reflecting surface, whereas for internal waves
they make the same angle with respect to the direction of {\em
gravity}.  When internal waves reflect off a sloping bottom, the
reflected wave has the same frequency $\omega$ as the incident wave.  But
because the wave frequency is related to the direction of
propagation by $\omega=N\sin \beta$  preservation of $\omega$ implies
preservation of
the angle $\beta$ ($N$ is the buoyancy frequency and, as shown in
figure~\ref{geometry}, $\beta$ the angle between the group velocity and the
horizontal). This reflection law leads to a concentration of the
energy density into a more narrow ray tube upon reflection. A fraction
of the energy of the incident wave is radiated away as the reflected
ray, but a part of the energy is lost due to turbulent dissipation, the
rest being converted into irreversible mixing which accounts for the
vertical transport of mass and matter.

Probably the most effective situation for boundary mixing arises when
an oncoming wave reflects from a bottom slope which nearly matches the
angle of wave propagation.  At this critical angle, the analytic theory of
internal waves reflecting from a uniformly sloping bottom (Phillips 1966)
predicts that the reflected wave has infinite amplitude and infinitesimal
wavelength. These unphysical results signal the failure of the idealizations
(for example, linear waves and inviscid fluid) made by \cite{phillips}.

\cite{sandstrom66} reported observational evidence of energy
enhancement at a particular frequency.  Later Eriksen (1982 and
1985) presented data showing energy and shear enhancement near the
critical frequency at a few mooring sites.  Eriksen interpreted
deviations from a linear and inviscid theory as evidence for
dissipation through shear instability and nonlinear interaction.  The
inviscid nonlinear case was theoretically considered by \cite{thorpe}
who showed that singularities may occur at other frequencies when a
finite amplitude incident wave interacts resonantly with its own phase
locked reflected wave.  \cite{gilbert} has also studied this
phenomenon on the continental rise and slope off Nova Scotia. Energy
enhancement at the critical frequency was significant at the 95\%
level for 8 of the 30 tests performed even though the overall
concavity of the slope should have slightly inhibited the phenomenon.
Finally, \cite{eriksen98} has recently reported striking observations
made on the steep flank of a tall North Pacific Ocean
seamount. Eriksen found a dramatic departures from the quasi-universal
Garett-Munk spectral model near the bottom in a frequency band
centered on the frequency for which ray and bottom slopes match.

\subsection{Laboratory and numerical experiments}
\label{Laboratory numericalexperiments}
Following the first observational studies, \cite{sandstrom66}
performed a laboratory experiment that clearly demonstrated the
amplification  that results from internal wave reflection
off a sloping bottom.  Then, \cite{cacchione} showed that at the
critical angle, the strong shearing motion becomes unstable and results
in the  formation of  a serie of periodic vortices.  Overturning of
these vortices produces mixed fluid that propagates into the interior as
regularly spaced layers all along the slope.

\cite{thorpehaines} reported evidence of the three dimensional
structure of the boundary layer but they were unable to reproduce the
formation of the vortex array seen by \cite{cacchione}. The absence of
vortices were also noted by \cite{iveynokes} when they studied the
mixing efficiency in the case of the breaking of monochromatic
incident wavefield uniformly distributed over the sloping bed.
\cite{taylor} was particularly interested in the decaying turbulence
and the mixing generated in the boundary layer.  \cite{imberger97}
considered the case of a small ray's width compared to the bed length.

\cite{iveynokes}, \cite{imberger97} and \cite{macphee} have also  shown the
formation of filaments by instabilities as waves approach the critical
frequency by the action of the incident and reflected waves.  Exactly
at the critical angle, instead of producing fine structured filaments,
the waves produced turbulent vortices at the boundary. These vortices
apparently mix boundary fluid which is presumably expelled along the
isopycnal corresponding to the new density of the mixed fluid.  As
suggested by \cite{Caldwell}, these intrusive layers could also
explained the presence of a highly ``stepped'' temperature profile as
the steep slope is approached at Lake Tahoe, California.

\begin{table}
  \begin{center}
  \begin{tabular}{@{}lcccccccccc@{}}
                     & $\gamma$  & $\beta$     & $N\sin\beta$& $h$   &$a_0$
  &$\tilde{\Rey}$ &\Rey & $a$
&$\varepsilon$&$\varrho$\\[3pt]
\cite{cacchione}   & 30$^{\circ}$& 30$^{\circ}$& 0.48    & 40  &0.4
  &2  &5400& 0.008 & 0.24&0.45\\
\cite{thorpehaines}& 20$^{\circ}$& 21$^{\circ}$& 0.65    & 28  &2.1
  &28 &4500& 0.037 & 0.32&0.13\\
\cite{iveynokes}   & 30$^{\circ}$& 30$^{\circ}$& 0.30    & 40  &4.6
  &139&3600& 0.085& 0.53&0.05\\
 \null\hskip1.2truecm ---   & --& --& --& --                   &3.7
  &89 &--& 0.068& 0.49&0.07\\
\cite{taylor}     & 20$^{\circ}$& 20.1$^{\circ}$& 0.16    & 53  &3.2
  & 14&4000&0.028&0.29&0.18 \\
 \null\hskip1.2truecm ---  & --& --& --    & --  &4.6  & 30&--&0.040&0.32&0.13 \\
\cite{imberger97}  & 38$^{\circ}$& 38$^{\circ}$& 0.36    & 40  &3.1
  &144&8200& 0.052& 0.61&0.03
 \end{tabular}
  \end{center}
\caption{A summary of experimental parameters. $\gamma$  is the
angle between the bottom slope  and
the horizontal. $N\sin\beta$ the frequency in s$^{-1}$; $h$ is the depth of
water in centimeters. $a_0$ is the amplitude of the motion of the paddle in
centimeters; $\tilde{\Rey}$=$\zeta^2N\sin\beta/\nu$ is the Reynolds number
defined by~Thorpe \& Haines (1987) and used
by Ivey \& Nokes (1989).
$\Rey=N\sin\beta/\nu K^2$ is the Reynolds number defined
  by Slinn \& Riley (1996)
and us. $a$ is the nondimensional parameter
 defined in (2.12). $\varepsilon=(a\tan\alpha)^{1/3}$ is the
small parameter  used in our  expansion and $\varrho$ is dissipative
parameter defined  in (3.8).}
\label{table1}
\end{table}

On the numerical side, \cite{slinnriley} have shown the creation of a
thermal front moving upslope at the phase speed of the oncoming wave
in the turbulent boundary layer. For a steep front, the thermal front
resembles a turbulent bore exhibiting nearly continuous localized
mixing, whereas for shallower slope, the mixing is observed across the
breadth of the domain and is temporally periodic. The internal wave
field continuously pumps fresh stratified fluid into the mixed layer,
while simultaneously extracting the mixed boundary fluid.
\cite{slinriley98} have pursued their published work and have
recently reported very detailed results on the effects of the
slope angle on the turbulent layer thickness, mixing period and
mixing efficiency.
An advantage of the numerical approach is that it is
possible to perform experiments over the shallow bottom slopes
that are more typical of oceanic conditions.

Other recent numerical work of \cite{javam} showed that for
near critical waves, the instabilities were triggered at the
bed, while for supercritical waves the instabilities develop  away from
the bed. \cite{javam} also showed the nonlinear creation of harmonics.

\subsection{Questions and strategy}

Our goals in this paper are to
understand the role of the nonlinearity in the reflection process, and to
to characterize the instabilities responsible
for the transition to turbulence near the slope. 
We also emphasize the role of
transience and  dissipation in healing the singularity which occurs when
the reflection is critical.

The paper is organized into four different sections.  The formulation
of the problem is described first in Section~\ref{analysis}.  We also
present the main effect of the reflection process in the outer region.
In Section~\ref{innerregion}, we focus our study to the reflection
problem in the boundary layer close to the slope. In
Section~\ref{inviscid}, we derive the explicit solution for the
inviscid case, whereas the viscous effects are presented in
Section~\ref{viscous}.  Finally, Section~\ref{conclusion} contains the
summary and conclusions.

\section{Analysis}
\label{analysis}

\subsection{Formulation in the coordinate system of the slope}

We will consider a two-dimensional, nonrotating, incompressible Boussinesq
fluid, with constant Brunt-V\"ais\"al\"a frequency $N$ and  a uniformly
sloping bed (angle of inclination is $\gamma$) shown in
figure~\ref{geometry}. We do not make the approximation of small
inclination. In lakes, bed slopes are in the range 2$^{\circ}$-20$^{\circ}$,
whereas in the ocean the {\it r.m.s.} slope of the sea bed is roughly
4$^{\circ}$. In the vicinity of seamounts slopes can be considerably higher
in the range 6$^{\circ}$-26$^{\circ}$ (see~\cite{imberger97} and references
therein). And most experiments use even larger angles of inclination.

The incident wave is a nearly monochromatic group of
internal gravity waves.  The angle between the incident group velocity and
the horizontal is $\beta$ and thus the carrier frequency of the group is
$ N \sin \beta$.

Our analysis employs a slope--oriented coordinate system in which $x$
is the distance along the slope and $z$ is the distance normal to the
slope. In terms of these tilted coordinates the  stratification of the
Boussinesq fluid is
\begin{equation}
\rho=\rho_0 \left[ 1 - g^{-1} N^2 (z \cos\gamma + x \sin\gamma)
- g^{-1} b\right],
\end{equation}
where $b(x,z,t)$ is the buoyancy perturbation of the resting
stratification.

Denoting by $(u,w)$ the components of the velocity field,
the  equations of  motion are
\begin{subeqnarray}
{Du \over Dt} - \sin \gamma \,b + p_x &=& \nu \nabla^2 u,\\
{Dw \over Dt} - \cos \gamma \,b + p_z &=& \nu \nabla^2 w,\\
{Db \over Dt} + N^2 \sin \gamma \,u + N^2 \cos \gamma \,w &=& \kappa
\nabla^2 b,\\
u_x+w_z&=&0,\label{equationsbeginnig}
\end{subeqnarray}
where $D/Dt=\partial_t + u\partial_x + w \partial_z$ is the convective
derivative.

The boundary conditions at $z=0$ are
\refstepcounter{equation}
$$
u=w=0, \quad N^2 \cos \gamma +
b_z=0.\eqno{(\theequation{\itl{a}-\itl{c}})}
\label{boundcond}
$$
The condition in (\ref{boundcond}c) is that there is no diffusive flux
of buoyancy through the slope.  \cite{wunsch70} and \cite{phillips70}
have shown that this condition at a sloping boundary  produces a flow
near the wall in a stably stratified fluid. Under laboratory conditions
using dissolved salt this flow is very small (but detectable). This no flux
condition will cause some complications later.

\subsection{ The incident wave and nondimensional variables}

\begin{figure}
\psfig{figure=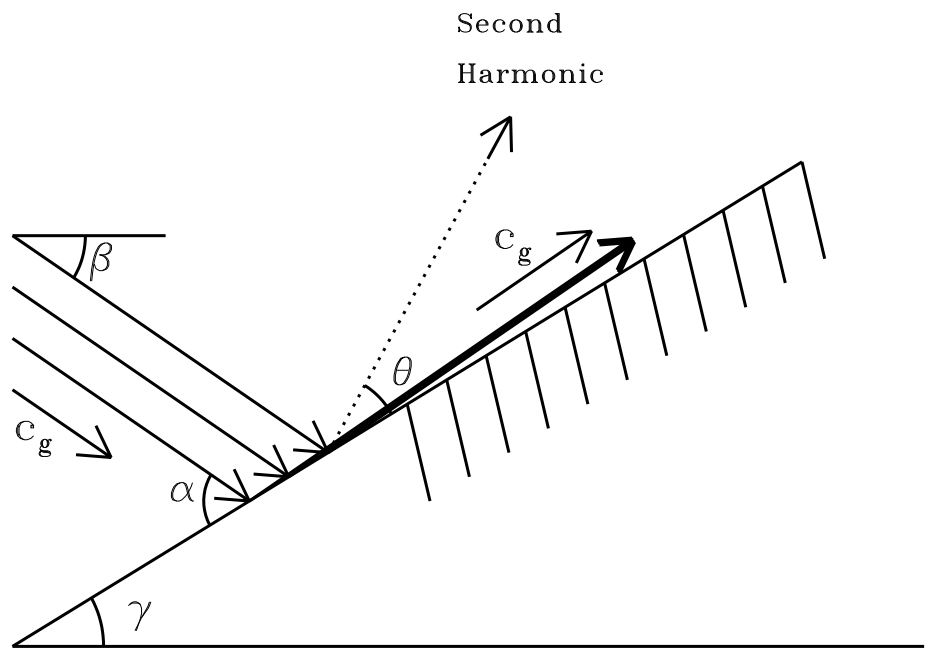,height=15truecm,width=15truecm}
\null\vskip -2truecm\null\hskip3truecm
\caption{Schematic view of the reflection process   for
$\gamma \approx \beta$.  The reflected wave
is alongslope, whereas the nonlinearly
reflected second harmonic, represented  with the dotted
line, makes an angle $\theta$ with the
slope.}
\label{schematicreflexion}
\end{figure}

The situation we envisage is shown in
figure~\ref{schematicreflexion}. The incident wave nearly satisfies
the critical condition that $\gamma=\beta$. Consequently the group
velocity of the reflected wave makes a very shallow angle with the
slope. The figure also anticipates some aspects of our analysis: when
the reflection is nearly critical the advective term  becomes
important in a region near the slope.  This quadratic nonlinearity
results in the production of a second harmonic. This nonlinearly
generated wave   can then radiate away from the
slope (see Thorpe 1987). The angle $\theta$ with which the second harmonic
leaves the slope is not necessarily shallow.

Using  the  convention
\begin{equation}
u=-\psi_z, \qquad w=\psi_x,
\end{equation}
for the  streamfunction, we consider an `incident wave train'
\begin{equation}
[\psi, b] \approx [1,NK] A(t - s z) \cos(K\sin\alpha\ x +  K\cos\alpha \
z - N\sin\beta t)\label{incidentwave}
\end{equation}
impinging on the slope. We have introduced the wave number $K$ and,
as suggested by figure~\ref{geometry},
\begin{equation}
\alpha \equiv \beta + \gamma.
\label{defalpha}
\end{equation}
The envelope, $A$ in (\ref{incidentwave}), changes
slowly relative to the space and time scales of the carrier wave;
$s^{-1}$ is the vertical group velocity of this slowly modulated
incident wave. Thus, near the slope, the incident wave switches on
slowly with a prescribed amplitude
$A(t)$.

We now nondimensionalize (\ref{equationsbeginnig})
 using the scales suggested by the
incident wave in (\ref{incidentwave}). The envelope can be written as
\begin{equation}
A= \psimax \hat A,
\end{equation}
where $\psimax$ is the maximum amplitude of the streamfunction and so
${max}(\hat A)=1$. The nondimensional variables are
\begin{subeqnarray}
\gdef\thesubequation{\theequation{\it a-c}}
[\hat x,\hat z]\equiv K[\sin\alpha\, x, \cos \alpha\, z],
 \qquad \hat t \equiv N\sin\beta\, t,\\
\gdef\thesubequation{\theequation{\it d-h}}
[\psi,u,w,b,p] \equiv \psimax [\hat \psi,\, K\cos\alpha\ \hat u,\,
 K\sin\alpha\ \hat w,\, K N\hat b,\, N\hat p].
\end{subeqnarray}
\returnthesubequation
Using the variables above, the nondimensional equations of motion are
\begin{subeqnarray}
{Du \over Dt} +{\tan\alpha\over \sin \beta} p_x -
{\sin\gamma\over\sin\beta\cos\alpha}\ b&=&{1\over \Rey} \nabla^2 u,\\
{Dw \over Dt}  +{\cot\alpha\over \sin \beta} p_z -
{\cos\gamma\over\sin\beta\sin\alpha}\ b&=&{1\over \Rey}\nabla^2 w,\\
{Db \over Dt} + {\sin \gamma\cos\alpha\over\sin\beta} \,u
              + {\cos \gamma\sin\alpha\over\sin\beta} \,w &=& {1\over\Pen}
\nabla^2 b,\\
u_x+w_z&=&0.
\label{nondimesionequations}
\end{subeqnarray}
The differential operators in (\ref{nondimesionequations}) are
\begin{subeqnarray}
\nabla^2 &=&\sin^2\!\alpha \, \partial_x^2 + \cos^2\!\alpha
\, \partial_z^2,\\
{D\over Dt} &=&\partial_t
+a\left(u\partial_x+w\partial_z\right).
\label{defnonlinearity}
\end{subeqnarray}
In (\ref{nondimesionequations}),  we have dropped the ``hats''
which would otherwise decorate the nondimensional variables. The
dimensionless dissipation parameters are the Reynolds and the P\'eclet
 numbers
defined as follows:
\begin{subeqnarray}
\gdef\thesubequation{\theequation{\it a,b}}
\Rey={N\sin\beta\over\nu K^2} \qquad
\mbox{and}\qquad
\Pen={N\sin\beta\over\kappa K^2}.
\label{defReynolds}
\end{subeqnarray}
\returnthesubequation
The other nondimensional parameter in (\ref{defnonlinearity}b)
 is a measure of the nonlinearity:
\begin{equation}
a \equiv { K^2\sin2\alpha\over 2N\sin\beta}\psimax .
\label{defdea}
\end{equation}
\cite{thorpe} gives a useful physical interpretation of the amplitude
parameter $a$: the maximum  slope of the isopycnals in the incident
wave $(2.9)$ is $a \tan \beta /(1-a)$ if $a<1$ or infinite if $a>1$. In
the sequel we will deal exclusively with the weakly nonlinear case in
which $a$ is small. The parameter $a$ is  also related to the internal Froude
number $\Fr\equiv\max(u_z/N)$ by
\begin{equation}
\Fr=a\,{ \sin2\beta\cos\beta \over \sin2\alpha}.
\end{equation}
Some authors prefer to use the minimal Richardson number
$\Ri\equiv\Fr^{-2}$. Thus, with the appropriate geometric factors,
$a$ is simply related to the other measures of nonlinearity used in the
literature.

In this nondimensional and slope-oriented coordinate system, the
dispersion relation of an inviscid linear internal wave (with all fields
proportional to $\exp [\ii (kx +  mz -  \omega t)]$) is
\begin{equation}
\omega =\pm{k \sin\alpha\cos \gamma - m \cos\alpha\sin \gamma
\over\sin\beta \sqrt{(k\sin\alpha)^2 +(m\cos\alpha)^2}},
\label{dispersionrelation}
\end{equation}
with  the corresponding group velocity
\begin{equation}
{\bmi c}_g = \left({\partial\omega\over \partial k},{\partial\omega\over
\partial m}\right)
=\pm{\sin2\alpha\left(m\cos\alpha\cos\gamma+k\sin\alpha\sin\gamma\right)
\over2 \sin\beta\left[(k\sin\alpha)^2
+(m\cos\alpha)^2\right]^{\frac{3}{2}}}(m,-k).
\label{groupvel}
\end{equation}
One solution of the dispersion relation~(\ref{dispersionrelation}) is
$(\omega,k,m)=(1,1,1)$; this is the incident wave. In terms of the
nondimensional variables, the wave fields are
\begin{equation}
\left[ \psi, b,p\right] ={\frac{A}{2}}\, e^{\ii (x+z-t)}
\left[1,1,-\ii \cos\beta \right]+ \mbox{c.c}.
\label{nondimicw}
\end{equation}
where $A(t-sz)$ is the wave envelope and
 $s^{-1}=-\sin2\alpha/2\tan\beta$ is  the group velocity in the
direction normal to the slope (put  $k=m=1$ in (\ref{groupvel})).

\subsection{The  `near-critical' approximation}

We will develop a reductive approximation which is based on taking a
distinguished limit in which $a$, $\beta-\gamma$, $\Rey^{-1}$ and $\Pen^{-1}$
are all small. To motivate our scaling assumptions,  we recall
the classical  solution of the reflection problem given by \cite{phillips}.
The no--flux boundary condition that $w=0$ at $z=0$ is satisfied by
superposing a reflected plane wave on top of the incident wave
in~(\ref{nondimicw}).  In terms of nondimensional variables, the vertical
wavenumber of the reflected wave is
\begin{equation}
m = {\tan (\gamma+\beta) \over \tan (\gamma-\beta) }.
\label{diverge}
\end{equation}
The {\em critical reflection} condition is that $\beta=\gamma$ and then,
according to~(\ref{diverge}), the wavenumber of the reflected wave is
infinite.  When $\gamma-\beta$ is small, $m$ is large and the
reflected wave has a much smaller length scale than that of the incident
wave. This singularity, or near singularity, signals that the
assumptions implicit in the Phillips' solution (stationarity, linearity
and absence of dissipation) fail when the incidence is near critical.

We define the  parameter $\varepsilon$ by\footnote{Because $a \propto
\cos\alpha$, $\varepsilon$ is
bounded as  $\alpha\to\pi/2$.}
\begin{equation}
\varepsilon\equiv(a\tan\alpha)^{\frac{1}{3}}.
\end{equation}
The reduction is based on the assumption that $\varepsilon$ is small. As a
measure of the degree to which the slope departs from the critical condition
$\beta =\gamma$, we introduce $\sigma$ defined
\refstepcounter{equation}
$$
\beta ={\alpha\over 2}-{\sigma\varepsilon^2 \over 2} , \qquad
\gamma={\alpha\over 2}+{\sigma\varepsilon^2 \over 2}.
\eqno{(\theequation{\itl{a},\itl{b}})}\label{defepsilon}
$$
The sign of  $\sigma$ determines if the reflection is supercritical
($\sigma<0$, upslope reflection) or subcritical ($\sigma>0$,
 downslope reflection).  $\sigma=0$ corresponds to precisely critical
reflection.

From (\ref{diverge}) and (\ref{defepsilon}) it follows that when
$\varepsilon \ll 1$,
\begin{equation}
m \approx {\tan\alpha \over \sigma\varepsilon^2}.\label{mapprox}
\end{equation}
Thus the `near-critical' condition in (\ref{defepsilon}) ensures that the
vertical wavenumber of the reflected wave is much greater than that of
the incident wave.

Using (\ref{groupvel}), one can also show that the group velocity
of the reflected wave in the near-critical case is
\begin{equation}
{\bmi c}_g\approx {\tan2\beta\over\tan\beta}{1\over m^2}[-m,1].
\label{cgcrit}
\end{equation}
Using~(\ref{mapprox}), one sees that the $z$-component
of the group velocity in (\ref{cgcrit}) is of order $\varepsilon^4$.
Thus if  $\varepsilon \ll 1$ it might take an  impractically long time to
establish the solution given by \cite{phillips}.

The theory employs a matched asymptotic expansion in which the
incident wave in~(\ref{nondimicw}) is in the {\it outer region} and the
reflected disturbance is largely confined to an {\it inner region},
which is essentially a boundary layer close to the slope.

  From~(\ref{mapprox}), we can anticipate that as $\varepsilon \to 0$ a
useful inner coordinate is likely to be
$\xi=\varepsilon^{-2}\tan\alpha\ z$. The fields of the reflected wave are
\begin{eqnarray}
\left[ \psi, b,p\right] ={\frac{A}{2}}\, e^{\ii (x+mz-t)}
\left[1,{\sin(\beta+\gamma)\over \sin(\beta-\gamma)},
\ii \cos\beta \right]+ \mbox{c.c}.
\label{scalforb}
\end{eqnarray}
and this suggests the introduction   of   the more appropriate variables
$B=b{\varepsilon^2/\sin\alpha}$ and $P=p/\cos{\beta}$.

Using (\ref{cgcrit}),  we can also estimate the  time for the reflected
wave to travel through a distance on the order of its own wavelength. One
finds that this transit time scales as $\varepsilon^2$ which motivates the
introduction of a {\it slow timescale} $t_2\sim\varepsilon^2 t$.

The other scaling assumptions  are that
\refstepcounter{equation}$$
\Rey ={\sin^2\alpha\over \varepsilon^{6} \nu_6} \qquad\mbox{and}\qquad
\Pen ={\sin^2\alpha\over \varepsilon^{6} \kappa_6}.
\eqno{(\theequation{\itl{a}-\itl{b}})}\label{scaling}
$$
The main justification for the choices above is {\it a posteriori} ---
they work in the sense that the dissipative terms are comparable to
the others in the final amplitude equation.  Knowledge of
$\varepsilon$ is then the key to recovering a specific physical
situation.

\subsection{Nonlinearly reflected second harmonic}
\label{nlreflected}
Because of the quadratic terms in (\ref{defnonlinearity}b), one can
 anticipate, following \cite{thorpe}, the nonlinear creation of a
 second harmonic in the small [$z={\cal O}(\varepsilon^2)$] region in
 which the advective terms become important. However this second
 harmonic does not remain confined to the region in which $z={\cal
 O}(\varepsilon^2)$. The second harmonic can radiate into the outer
 region where it appears as a plane wave. A successful completion of
 the matching problem requires that one recognizes this possibility
 that the inner region drives the outer region through this nonlinear
 process.

We now assemble the fields which describe the second
harmonic in the outer region. A
modulated plane wave second harmonic  with upward
group velocity is
\begin{equation}
[\psi,b,p] ={\frac{1}{2}}
H  \, e^{\ii (2x + \tilde m  z - 2t)}
[1,\tilde b, \ii \tilde p]+\mbox{c.c.},
\label{secharmfields}
\end{equation}
where  $H(t-\tilde s z)$ is the envelope of the second harmonic
and $\tilde s^{-1}$ is the vertical group velocity. The
matching will determine $H$ in terms of the incident amplitude $A$. In
(\ref{secharmfields}), we use the symbol
 $\tilde{}$ to denote quantities associated
with the second harmonic. In the nearly critical case these quantities
are given to leading order by
\begin{subeqnarray}
\tilde m(\alpha)
&\equiv&-{2\over3}\cosec {\alpha\over2}
\tan\alpha\left[\cos{\alpha\over2} +
{2\left(2\cos\alpha-1\right)^{\frac{1}{2}}}\right],\\
 \tilde b(\alpha)&\equiv&1 +  \left( 1-{\tilde m\over 2} \right) \cos
        \alpha,\\
\tilde p(\alpha)&\equiv&\left(2\cos{\alpha}-1\right)^{\frac{1}{2}},\\
\tilde s(\alpha)&\equiv&-{\tan{\alpha}\over 2}
{\left(4+\tilde m^2\cot^2\alpha\right)^{\frac{3}{2}}
\over
2+{\tilde m}\cot\alpha\cot{\alpha\over2}}.
\end{subeqnarray}

\begin{figure}
\null\hskip 3.3truecm\psfig{figure=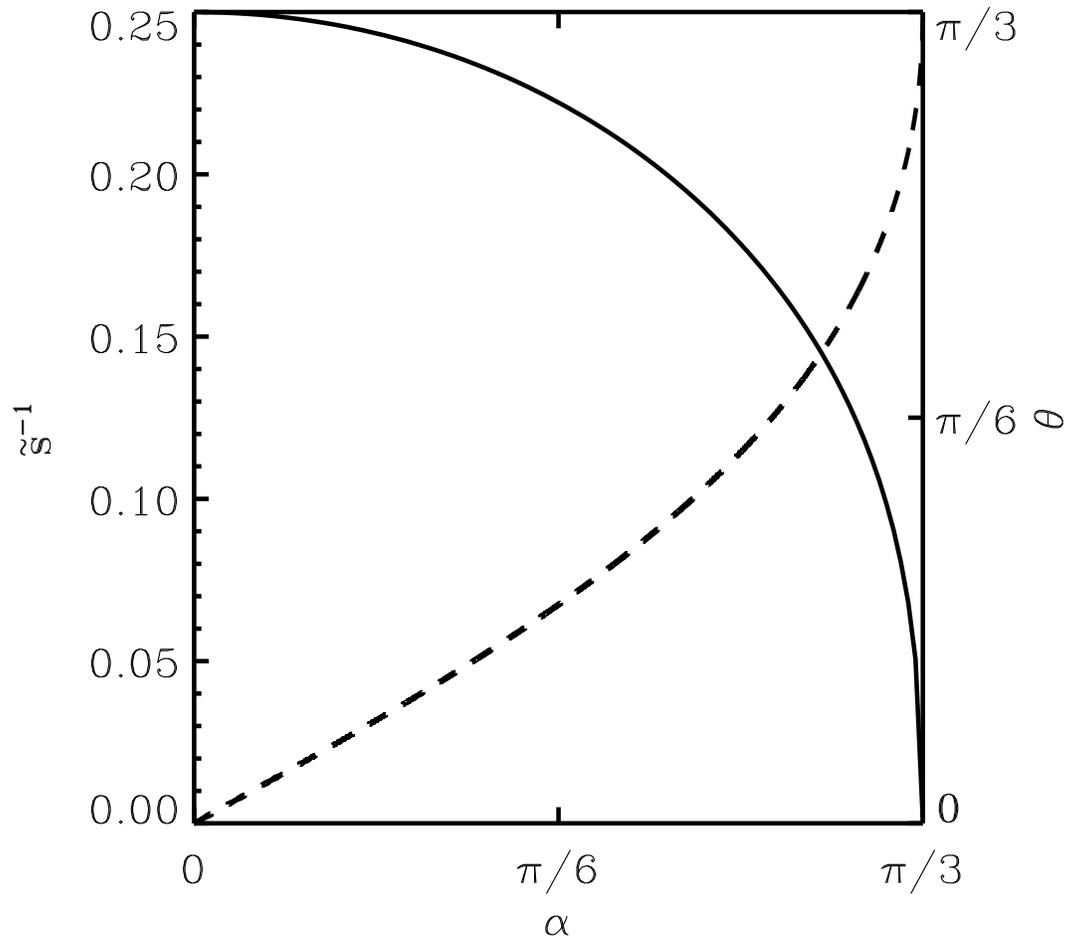,height=15truecm,width=15truecm}
\caption{$\tilde s^{-1}$ (solid line), the $z$-component of the
group velocity, and $\theta$ (dashed line)
angle of reflection  between the slope and
 the direction of propagation of  the
 second harmonic in the critical case.}
\label{stilde}
\end{figure}

\begin{figure}
\null\hskip3truecm\psfig{figure=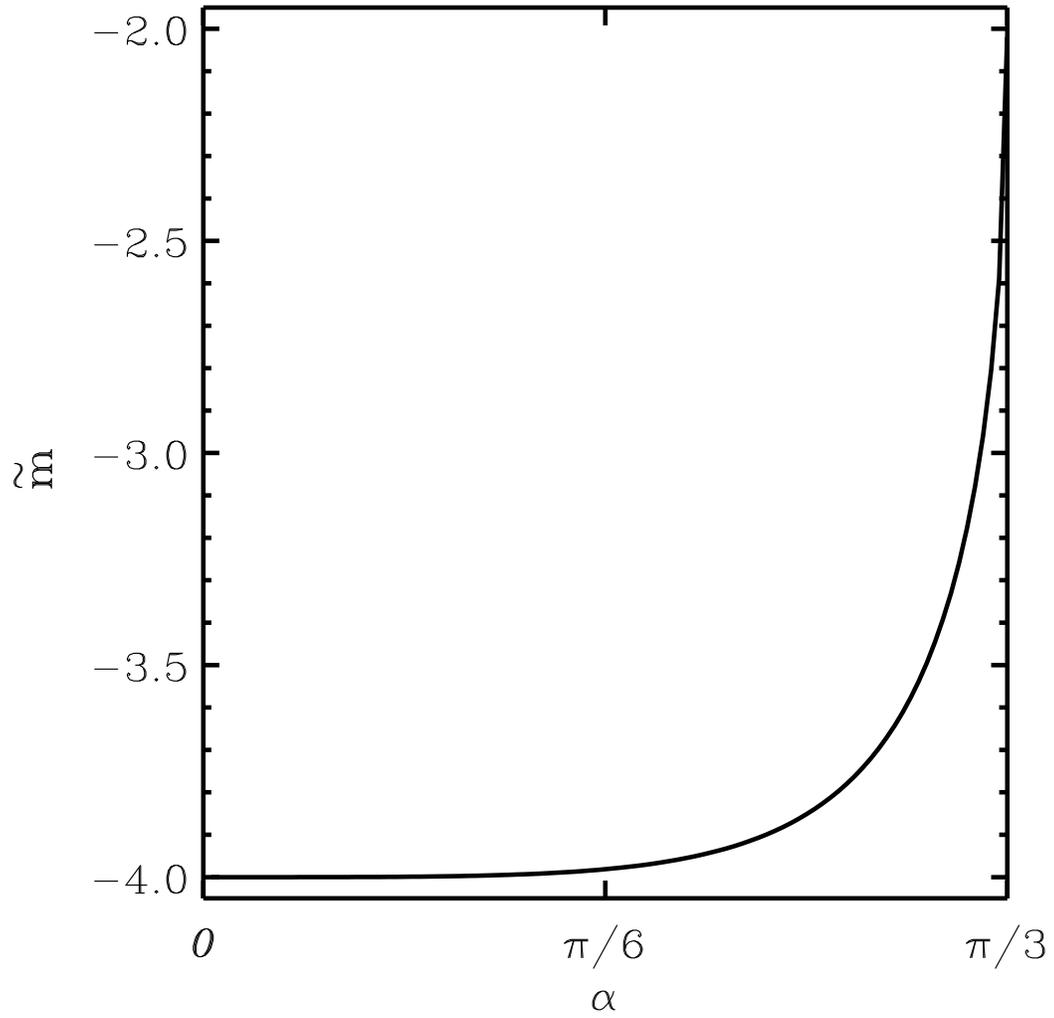,height=15truecm,width=15truecm}
\caption{ $\tilde m (\alpha)$, the $z$-wavenumber of the nonlinearly
reflected second harmonic, in the near-critical case.
When $\alpha>\pi/3$, $\tilde m (\alpha)$ is complex.}
\label{mtilde}
\end{figure}

If $\alpha>\pi/3$, the second harmonic is evanescent
\footnote{\cite{javam} showed the nonlinear creation of evanescent
harmonics. Apparently they never performed numerical experiments 
 in the case $\gamma<\pi/6$ when they should
propagate according to the present results.}
 in $z$, while for
$\alpha<\pi/3$ the second harmonic propagates away from the slope.
$\tilde m(\alpha)$ is plotted in figure \ref{mtilde}. The angle $\theta$
between the slope and the direction of propagation  of the second harmonic is
\begin{equation}
\theta=-\arctan\left({2\tan\alpha\over \tilde m(\alpha)
}\right),\hskip 2truecm (\alpha\le\pi/3).
\end{equation}
Both $\theta$ and the group velocity $\tilde s^{-1}$ are shown in figure
\ref{stilde}. Because both $\theta$ and the group velocity are nonzero the
second harmonic is not trapped in a boundary layer close to the slope.

\cite{slinriley98a} have reported that when $\gamma<\pi/6$, intrusive
layers forms near the slope. However, interestingly, in the case
$\gamma=\pi/6$ (i.e., the critical value above which the nonlinearly
reflected second harmonic is evanescent), there is a uniform
thickening of the dye layer along the slope.

\subsection{The outer region}
The first two orders of the outer solution are obtained
by taking a combination of the incident wave in (\ref{nondimicw})
and the second harmonic (\ref{secharmfields})
\begin{equation}
[\psi,b,p] ={\frac{1}{2}}\, A\, e^{\ii (x+z-t)}
\left[1,1,-\ii \cos\beta \right]+
{\varepsilon\over{2}}
H  \, e^{\ii (2x + \tilde m  z - 2t)}
[1,\tilde b, \ii \tilde p]+{\cal O}(\varepsilon^2)+\mbox{c.c.}.
\label{outercombin}
\end{equation}
The connection between $H(t-\tilde s z)$ and $A(t-sz)$ is determined by
the matching problem in the inner region.

\section{The inner region}
\label{innerregion}

In the inner region, we use a stretched
coordinate $\xi$ to describe  the reflected disturbance
close to the slope and a slow time scale $t_2$. Following our earlier
discussion in section 2.3, these  scales are:
\refstepcounter{equation}$$
\xi\equiv \varepsilon^{-2}\tan\alpha\ z
, \qquad t_2\equiv\mu\varepsilon^2 t,
\eqno{(\theequation{\itl{a},\itl{b}})}
\label{scal}
$$
where $\mu\equiv\cot(\alpha/2)$.
As suggested by (\ref{scalforb}), we  introduce
\refstepcounter{equation}$$
B \equiv {\varepsilon^2 b\over\sin\alpha}\qquad\mbox{and}\qquad
p \equiv \cos{\beta} \ P .
\eqno{(\theequation{\itl{a}-\itl{b}})}
$$
Finally,  it is convenient to define
\refstepcounter{equation}$$
u\equiv\varepsilon^{-2}\tan\alpha\  U ,\qquad w\equiv W,\qquad
\psi \equiv \Psi.
\eqno{(\theequation{\itl{a}-\itl{c}})}
$$
so that we have
\refstepcounter{equation}$$
U  =-\Psi_\xi\qquad\mbox{and}\qquad W=\Psi_x.
\eqno{(\theequation{\itl{a}-\itl{b}})}
$$
In the zone close to the slope, where $\xi={\cal O}(1)$, $B$ and $U$ are also
${\cal O}(1)$. That is, the buoyancy perturbation and the along-slope velocity
are ${\cal O}(\varepsilon^{-2})$ larger than in the incident wave.

In the inner region, using (\ref{scaling}),  (\ref{nondimesionequations})
becomes
\begin{subeqnarray}
{DU \over Dt} -B&=&\varepsilon^2\left[-\mu P_x +\mu\sigma B+\nu_6U_{\xi\xi}
\right]+{\cal O}(\varepsilon^3),\\
-P_\xi+B&=&\varepsilon^2\left[{\sigma\over 2\mu} B+{1\over
\mu}{\partial W\over \partial t }\right]+{\cal O}(\varepsilon^3),\\
U_x+W_\xi&=&0,\\
{DB \over Dt} + U &=& \varepsilon^2\left[-\mu\sigma U-\mu W+\kappa_6
B_{\xi\xi}\right]+{\cal O}(\varepsilon^3),
\label{equationsinnerregion}
\end{subeqnarray}
where the convective derivative is
\begin{equation}
{D \over Dt}  ={\partial_t
}+\mu\varepsilon^2\partial_{t_2}+\varepsilon
\left(U\partial_x+W\partial_\xi\right).
\end{equation}

Using  the complex variable $S=\Psi+\ii P$, (\ref{equationsinnerregion})
can be written compactly as
\begin{eqnarray}
S_{\xi t} -\ii S_\xi=-
\varepsilon J(\Psi,S_\xi)&&+
\varepsilon^2\Bigl[-\mu S_{\xi t_2}+\ii\mu\left(\sigma S_\xi-S_x\right)
-{\ii\over \mu}\left(W_{tt}-\ii W_t\right) \nonumber\\
&&-{\ii \sigma\over 2\mu}
\left(B_{t}-\ii
B\right)+\mu\varrho^2 S_{\xi\xi\xi}+\mu\zeta S_{\xi\xi\xi}^*\Bigr]
+{\cal O}(\varepsilon^3),\qquad \label{finalequation}
\end{eqnarray}
where
\refstepcounter{equation}$$
\varrho^2={\nu_6+\kappa_6\over2\mu},\qquad
\zeta={\nu_6-\kappa_6\over2\mu},
\eqno{(\theequation{\itl{a},\itl{b}})},\label{defvarrho}
$$
and $J(a,b)$ is the Jacobian.

The equation above must be solved with the boundary conditions:
\begin{subeqnarray}
S(x,0,t,t_2)+S^*(x,0,t,t_2)=0,&&\qquad\qquad\mbox{(no normal flow)}\\
S_\xi(x,0,t,t_2)+S_\xi^*(x,0,t,t_2)=0,&&\qquad\qquad\mbox{(no slip)}\\
S_{\xi\xi}(x,0,t,t_2)-S_{\xi\xi}^*(x,0,t,t_2)=0&+\ {\cal O}(\varepsilon^4).
&\qquad\qquad\mbox{(no flux of buoyancy)}
\label{boundcondbis}
\end{subeqnarray}

By taking  the inner limit of
the right hand side  of (\ref{outercombin}), we obtain
the matching condition
\begin{eqnarray}
\lim_{\xi \to \infty} S =A(t_2) e^{\ii(x-t)}&+&{\varepsilon H}(t_2)
      {(1-\tilde p\sec\beta)\over 2}e^{-2\ii(t- x)}\nonumber\\
&+& {\varepsilon H}^*(t_2)
{(1+\tilde p^*\sec\beta)\over 2}e^{ 2\ii(t- x)}
+{\cal O}(\varepsilon^2).
\label{matchingconditions}
\end{eqnarray}

The weakly nonlinear
analysis proceeds by introducing not only slow space and time
scales but also seeking a solution of (\ref{finalequation}) in terms of
the regular perturbation expansions
\begin{equation}
S=S_0+\varepsilon S_1+\varepsilon^2S_2+ {\cal O}(\varepsilon^3).
\label{expansion}
\end{equation}

Substituting (\ref{expansion}) into (\ref{finalequation}) leads
to the following hierarchy:
\begin{subeqnarray}
\varepsilon^0:\qquad&\left(\partial_{t}-\ii\right)\partial_{\xi} S_0&=0,\\
\varepsilon^1:\qquad&\left(\partial_{t}-\ii\right)\partial_{\xi} S_1&=-
J(\Psi_0,S_{0\xi}),\\
\varepsilon^2:\qquad&\left(\partial_{t}-\ii\right)\partial_{\xi} S_2&=-
\left[J(\Psi_0,S_{1\xi})+J(\Psi_1,S_{0\xi})\right]+[LT]_0.\label{dorder}
\end{subeqnarray}
In (\ref{dorder}a), $[LT]_0$ means the linear term,
in the square bracket of the right hand side
 of (\ref{finalequation}),  evaluated with $S_0$.

\subsection{Leading order}

From (\ref{dorder}a), we obtain the  leading order solution
\begin{equation}
S_0=e^{-\ii( x-t)} \SS(\xi,t_2)+A e^{\ii(x-t)} -A^*
e^{-\ii(x-t)},\label{resultatforS_0}
\end{equation}
and the leading order streamfunction
\begin{equation}
\Psi_0={\frac{1}{2}}e^{\ii (t-x)} \SS(\xi,t_2)+\mbox{c.c.}.
\label{resultatforpsi_0}
\end{equation}
The evolution of $\SS(\xi,t_2)$ will be determined at higher order.
However, at this order,
the matching condition (\ref{matchingconditions}) is satisfied provided that
\begin{equation}
\lim_{\xi \to \infty} \SS= A^*,
\label{Sinfinity}
\end{equation}
whereas the no-flux and no-slip condition condition at $\xi=0$ requires that
\begin{equation}
\SS(0,t_2)=
\SS_{\xi}(0,t_2)=\SS_{\xi\xi}(0,t_2)=0.
\label{3conditions}
\end{equation}
Equation (\ref{resultatforS_0}) is not the most general solution
of (\ref{dorder}a). However, for simplicity we include only
$e^{-\ii x}$ harmonic which is required by the matching condition
to the incident wave.

\subsection{Order $\varepsilon$}

Equation (\ref{dorder}b) gives
\begin{equation}
S_{1\xi t}- \ii S_{1\xi}=-{1\over 2}
\left[{e^{2\ii t}} J(e^{-\ii x} \SS,e^{-\ii x} \SS_\xi)+
J(e^{\ii x} \SS^*,e^{-\ii x} \SS_\xi)\right],\label{tointegrate}
\end{equation}
leading to
\begin{equation}
S_1=\ii\R(x,t,t_2)+\, e^{2\ii (t-x)}
\left[{\frac{1}{2}}\SS\SS_\xi
- \int_{0}^\xi\SS_u^2\, \rmn{d} u\right]+\,
{\SS^*\SS_\xi\over 2}.
\label{resultatforS_1}
\end{equation}
We satisfy the no-normal flow condition (\ref{boundcondbis}a) by
requiring that $\R$ is real.  The no-slip condition
(\ref{boundcondbis}b) and the no-flux condition (\ref{boundcondbis}c)
are satisfied provided that $\SS$ satisfies these conditions. 
At this
order, the nonlinear effects produce a rectified and a second harmonic
waves.

Now we match (\ref{resultatforS_1}) with (\ref{matchingconditions}).
The most important result is that this matching condition
defines the envelope of
the nonlinearly reflected second harmonic:
\begin{equation}
H(t_2)=-
\displaystyle
\int_{0}^\infty \SS_\xi^{*2}\,\rmn{d} \xi  
\,.\label{Npoura2bis}
\end{equation}
Thus we have an expression for the amplitude
of the second harmonic in terms of the incident wave.
The matching tells also that $\R=He^{2\ii (x-t)}(1-\tilde
p\,\sec\beta)/2\ii +\mbox{c.c.}$ and that 
the rectified flow vanishes in the outer region because
of condition (\ref{Sinfinity}).

\subsection{Order $\varepsilon^2$}
At this order, we have (\ref{dorder}c). To avoid secular growth,
all the resonant terms on the right hand side must vanish.
This  condition determines the evolution equation for
$\SS(\xi,t_2)$.

Although it is not initially obvious, all nonlinear resonant
contributions cancel
\footnote{The underlying reason for the miraculous
cancellation of the resonant nonlinear terms is the
following special case of the Jacobi identity:\\
$J\left[\Q,J(\Q_\xi,\Q^*) \right]
+J\left[\Q^*,J(\Q,\Q_\xi) \right]
+J\left[J(\Q,\Q^*),\Q_\xi \right]=0$.} 
and the final evolution equation is
{\em linear}
\begin{equation}
\displaystyle
\SS_{t_2\xi}-\ii
\sigma\SS_\xi+\SS-\varrho^2\SS_{\xi\xi\xi}=A^*(t_2)
.\label{finalequa}
\end{equation}
Using (\ref{finalequa}), one can obtain an alternative expression for the
envelope of the second harmonic:
\begin{equation}
H_{t_2}+2\ii\sigma H=2\varrho^2\int_0^\infty
\SS_{\xi\xi}^2\,\rmn{d} \xi-A^{2}.
\label{equationforI}
\end{equation}

We emphasize that although there are no nonlinear terms in
(\ref{finalequa}), the nonlinearity is important for the generation of
the second harmonic and also for all the nonlinear contributions to
$S_1$ in (\ref{resultatforS_1}).
 And when we come to visualize the solution, these nonlinearly
forced components of the solution lead to a symmetry breaking.

Equation (\ref{finalequa}) is third order in space and we are imposing
four boundary conditions in (\ref{Sinfinity}) and (\ref{3conditions}).
Consequently, the problem is overspecified and we resolve this issue
by discarding the no buoyancy flux condition ${\cal S}_{\xi\xi}=0$.
This unsatisfactory point might be corrected by demanding that the
P\'eclet number in (\ref{defReynolds}b) be very large.  In this
circumstance, one would expect a very thin buoyancy diffusive layer in
which dynamics similar to that of \cite{wunsch70} and
\cite{phillips70} is important.

\section{The inviscid case}
\label{inviscid}

We first consider the  special case of (\ref{finalequa}) in which  the fluid
is inviscid  ($\varrho=0$). The solution of (\ref{finalequa}) which
satisfies the initial condition that $\SS(\xi,0)=0$ is
\begin{eqnarray}
{\cal S}(\xi,t_2) =\int_0^{t_2}A^*(t_2-\tau)e^{\ii\sigma \tau}
 \sqrt{\xi\over\tau}\
 J_1\left(2\sqrt{\xi \tau}\right)
\ \rmn{d} \tau,
\label{BillEq:1}
\end{eqnarray}
and (\ref{equationforI}) leads to the following
expression for the envelope of the second harmonic:
\begin{equation}
H(t_2)=-\int_0^{t_2}\!\!e^{2\ii \sigma(\tau- t_2)}\ A^2(\tau)\ \rmn{d}\tau.
\end{equation}

The amplitude $A(t_2)$ of the incident wave must be specified to
completely determine the solution.  As in all the experiments, we take
the simplest case in which $A(t_2)$ switches on
suddenly\footnote{Because $A$ depends only on the slow time $t_2$, the
sudden switch-on means that the incident wave achieves its ultimate
constant amplitude on a time scale which is slow relative to $t$ but
fast relative to $t_2$.} at $t_2=0$. That is, $A(t_2)=1$ when $t_2>0$.

\subsection{The critical case}
In the critical case $\sigma=0$ the integral in (\ref{BillEq:1}) can
be evaluated and \begin{equation} {\cal S}=1-J_0(2\sqrt{\xi t_2}),
\label{inviscidsolforf} \end{equation} where $J_0$ is the Bessel
function of the first kind of order 0.  Thus, in this critical case,
the solution has a `similarity' form in which the thickness of the
inner region is inversely proportional to time. That is, there is no
steady solution as $t\to \infty$. Instead, the disturbance near the
slope becomes strongly oscillatory as the undulations of $J_0$ are
intensified.

Using (\ref{inviscidsolforf}) we have:
\begin{subeqnarray}
[\Psi_0,U_0] &=&\cos(x-t)\left[1-J_0,
-{2t_2}{J_1\over\chi}\right],\\
\left[W_0,P_0,B_0\right]
&=&\sin(x-t)\left[J_0-1,1+J_0,-{2t_2}{J_1\over\chi}\right],\\
\Psi_1&=&t_2\left({J_1\over\chi}(1-J_0)+J_0^2+J_1^2-1\right)
\cos2(x-t)+t_2(1-J_0){J_1\over\chi},\\
U_1&=&2t_2^2\left({J_2(1-J_0)+J_1^2\over\chi^2}\right)
\cos2(x-t)+2t_2^2
\left({J_2\over\chi^2}(1-J_0)-{J_1^2\over\chi^2}\right),\hskip 1truecm\\
W_1&=&2t_2\left(1-J_0^2-J_1^2-{J_1\over \chi}(1-J_0)\right)
\sin2(x-t),\\
B_1&=&{2t_2^2}\left(
{J_2\over \chi^2}(1-J_0)+{J_1^2\over \chi^2}\right)
\sin2(x-t),
\label{critinte}
\end{subeqnarray}
where $\chi\equiv2\sqrt{\xi t_2}$ and $J_i=J_i(\chi)$. Notice that in this
critical case the along-slope velocity $U_0$ and the buoyancy perturbation
$B_0$ both grow linearly with time. This response is analogous 
to that of a resonantly forced oscillator.

\begin{figure}
\psfig{figure=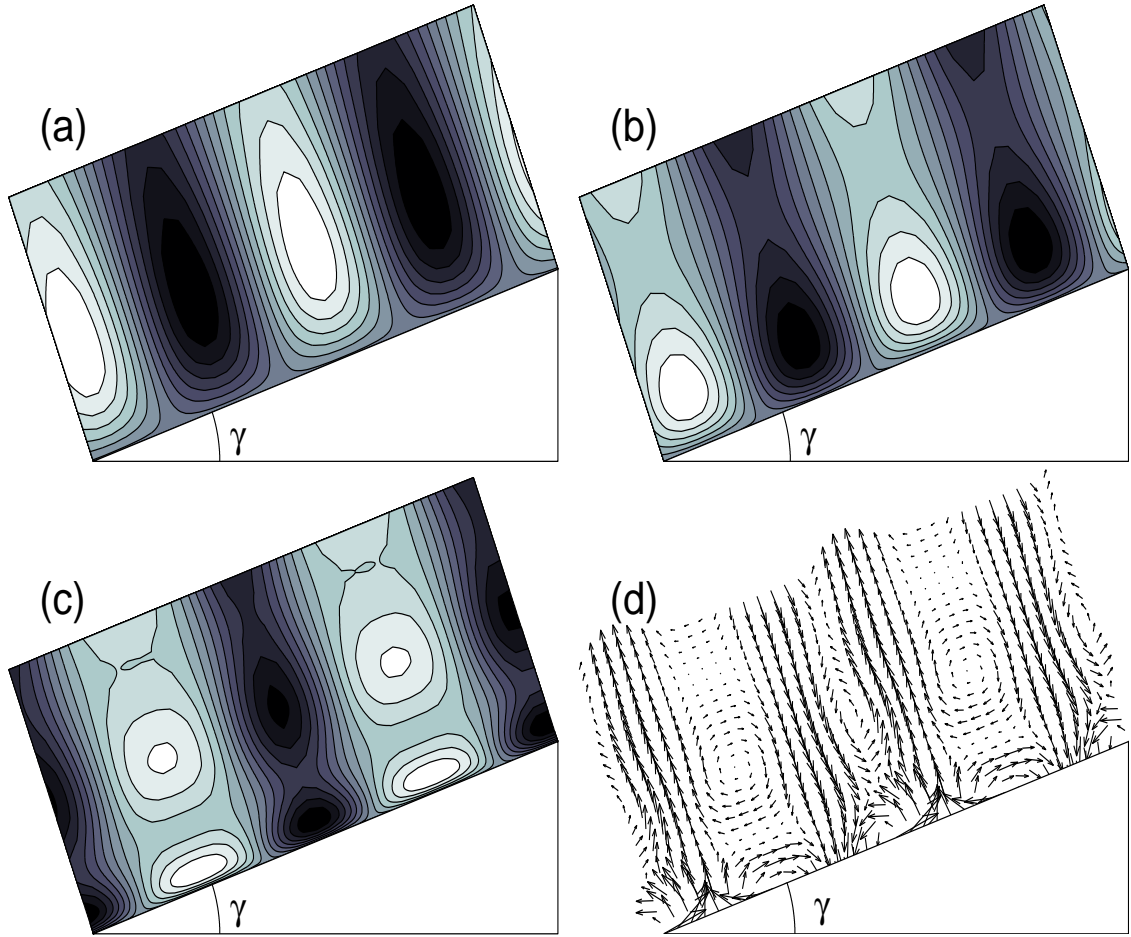,height=15truecm,width=15truecm}
\caption{Streamfunction for four different times
in the  critical case.
$N\sin\beta \ t={1\over2}$, 1 
 and 3 in the panels (a), (b) and (c). Panel (d) is an 
alternative ``velocity vector'' visualization of panel (c).
 $\varepsilon=0.3$, $\gamma=\beta= 20^{\circ}$.
In panel (d), 
there are two pairs of  counter-rotating vortices
immediately adjacent to the slope. The clockwise vortices
are slower and thinner than the counter-clockwise vortices.
This symmetry breaking is a result of the second
harmonic term such as $\Psi_1$ in ({\protect{\ref{critinte}}}).
The dimensions of the panel are $2\lambda/\sin\alpha$ in the
$x$-direction (i.e. two alongslope wavelengths)
by  $\lambda/\sin\alpha$ in the
$z$-direction.}
\label{psi4}
\end{figure}

The streamfunction is shown in figure~\ref{psi4}. At small times the
reflection process creates a regular array of counter-rotating
vortices. As time progresses, figure~\ref{psi4}(b) shows that the
scale of the vortices decreases.  Panels (c)
and (d) are both at $N\sin\beta\ t=3$. The ``velocity vector''
presentation in panel (d) more clearly displays the asymmetry of the
vortices which is the effect of the nonlinear terms in $\Psi_1$.

\begin{figure}
\null\hskip1truecm
\psfig{figure=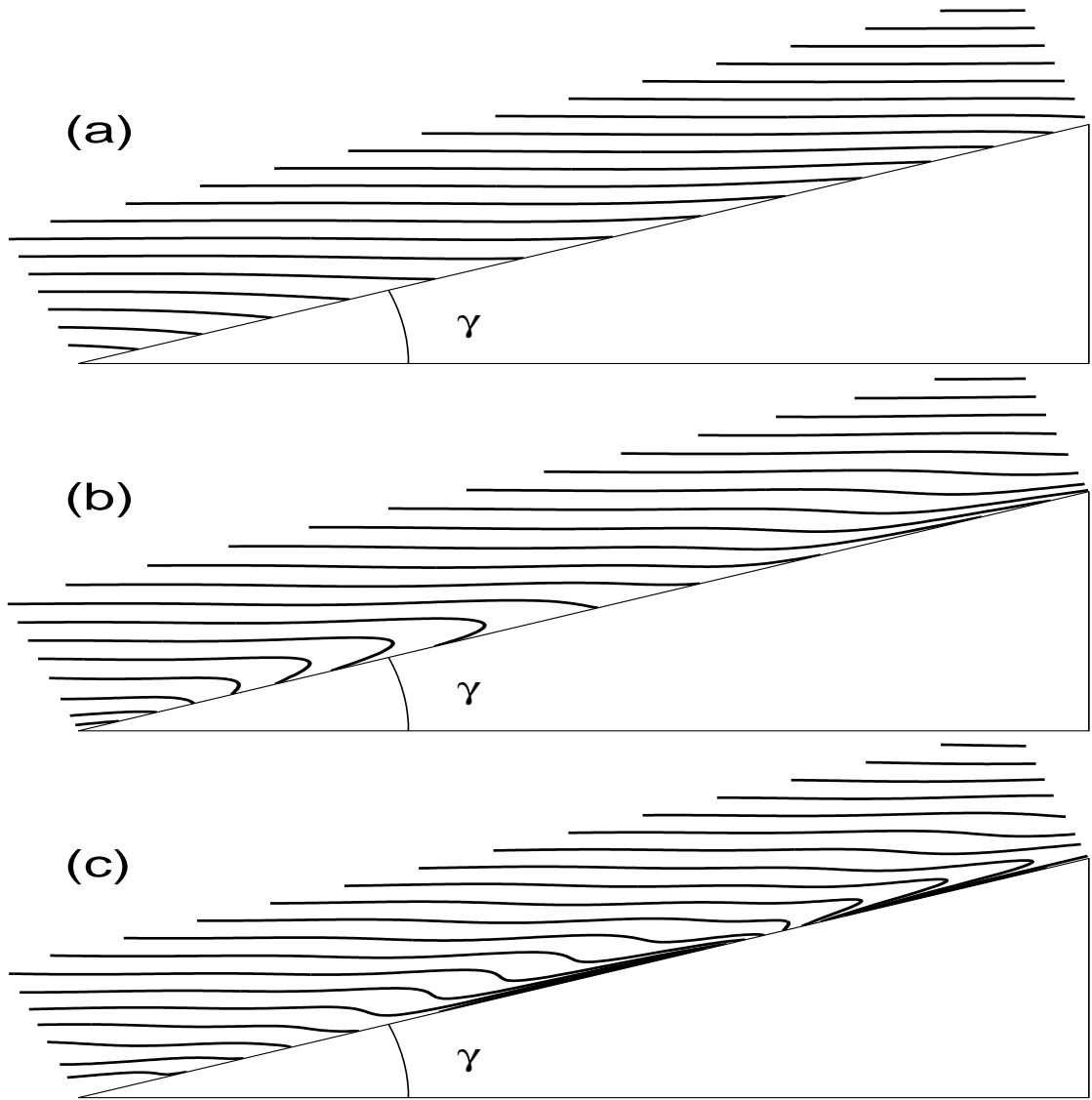,height=15truecm,width=15truecm}
\caption{Buoyancy field in the critical case
at $N\sin\beta \ t=$3, 10 and 20
in the panels (a), (b) and (c) respectively. $\varepsilon=0.3$;
$\beta=\gamma=20^{\circ}$.
The dimensions of the panel are $\lambda/\sin\alpha$ in the
$x$-direction by  $\lambda/5\sin\alpha$ in the
$z$-direction.
In the panels (b) and (c), some overturned regions are evident.}
\label{3buo}
\end{figure}

 Figure \ref{3buo} shows the distortion of  the isopycnals as the
oscillations amplify. In panel (a)  the
disturbance is very small and one sees  essentially the initial
background stratification.  In panel (b)  the
disturbance begins to  `fold-up' the isopycnals and, near the slope, this
process produces a region of static instability. In the panel (c) the
development of small scales in the isopycnal field is evident.

In this case, with $\sigma=0$, 
the amplitude of the  second harmonic is a linear function of time:
\begin{equation}
H(t_2)=-t_2=-{\varepsilon^2\over\tan\gamma}\ t,
\end{equation}
and so $\varepsilon \Psi_1$ becomes comparable to $\Psi_0$ when
$t_2={\cal O}(\varepsilon^{-1})$. For these reasons the expansion becomes
disordered when $t_2 = {\cal O}(\varepsilon^{-1})$ and the results above are
no longer reliable. However, as we show below, well before this 
 breakdown, the buoyancy becomes statically unstable. Thus the
expansion above strongly suggests that the next evolutionary stage is
characterized by the onset of turbulence triggered by overturning
instability.

\subsection{Overturning instability or stratified shear flow instability?}
\label{ConvectiveorShear}

One scenario for the transition to turbulence is that the growing
disturbance produces a statically unstable density field which then
overturns (we refer this as overturning instability). An alternative
is that the local Richardson number might fall below a critical value
while the density is still statically stable. In this second case the
overturns are produced by a rapidly growing secondary shear flow
instability.  Experimentally, \cite{thorpehaines} have reported that
the overturning instability is very likely to be convectively driven.
On the other hand, \cite{slinriley98a} identified a shear instability
mechanism; but they noted that the Reynolds number of their numerical
simulations was matched to experimental values by forcing larger
amplitude waves in a more viscous fluid. Consequently, the Richardson
number in these simulations was relatively low compared to those of
laboratory experiments.  However, all these claims must be viewed
cautiously because both instabilities are intrinsically related in a
stratified flow.

Using our analytic results we can make a rough assessment of these two
possibilities.  Taking into account the background linear buoyancy,
and the first order correction, we can calculate the overturning time
$t_o$ which is the time at which negative vertical buoyancy gradients
first occur.  In dimensional variables: \begin{equation}
t_o={\sqrt{8\tan\gamma}\over N \sin 2\gamma} \
\varepsilon^{-{3\over2}}.  \label{defto} \end{equation} We can also
calculate the time $t_s$ at which the minimum Richardson number first
falls below $1/4$: \begin{equation} t_s=\sqrt{2 \over \cos \gamma} t_o
>t_o. \label{defts} \end{equation} This analytical calculation also shows that both
of the unstable conditions above occur first at the wall (see also
\cite{javam}).

\begin{figure}
\null\hskip
3truecm\psfig{figure=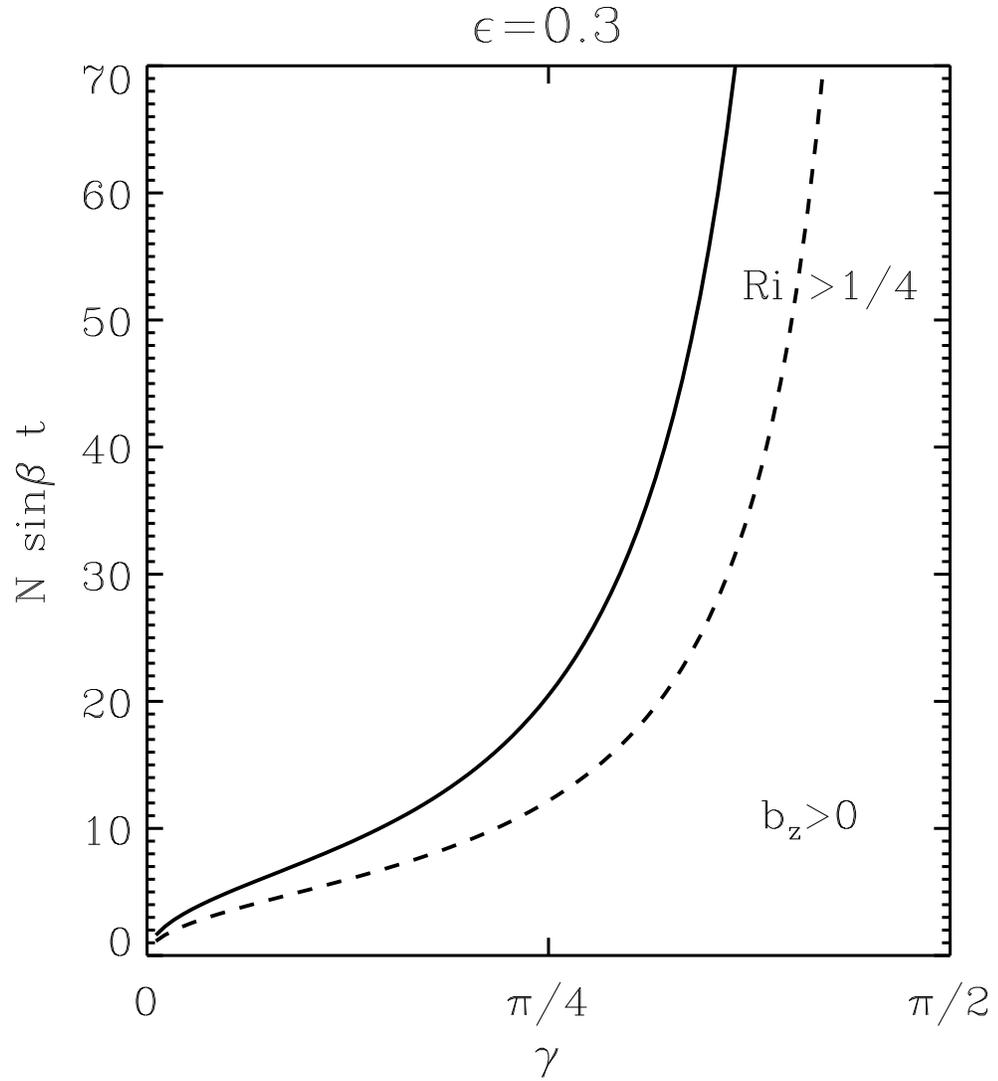,height=15truecm,width=15truecm}
\caption{$N\sin\beta \ t_o$ (dashed line)
and $N\sin\beta \ t_s$ (solid line) as a function of the slope angle
$\gamma$.}
\label{overturnshear}
\end{figure}

Because $t_o<t_s$ we can argue that the convectively driven
overturning instability should appear first. In the typical
experimental case $\varepsilon=0.3$, for example, both critical times
are plotted as a function of the slope angle $\gamma$ in
figure~\ref{overturnshear}.  It is interesting to note that
$N\sin\beta\ t_o$ is an increasing function of the slope angle
$\gamma$.

However, let us note that the domain of validity of the
Miles-Howard theorem (stability if $\Ri>1/4$) applies to steady,
parallel shear flows, whereas the present flow is unsteady and
nonparallel. 
In addition, Eq.~(\ref{defts}) was derived using the usual
definition of the Richardson number $\Ri=N^2/U_z^2$
(see, for example, \cite{kundu}).
An alternative definition could be 
\begin{equation}
{\tilde{\Ri}}=-{ (g/\rho_0){d\rho\over dz}\over
{U_z}^2},
\label{Rireferee} 
\end{equation}
where the numerator takes into account not only the background
buoyancy, but also the perturbative part of the buoyancy.  It is clear
from Eq.~(\ref{Rireferee}), that ${\tilde{\Ri}}$ would reach 1/4
before zero, and therefore, before negative vertical buoyancy
gradients first occur.  So, the whole discussion simply indicate that
there are plausible reasons to expect that the solution we have found
will become unstable, and ultimately turbulent, and we have {\em
three} criteria for instability; the static instability time is
sandwiched between the ${\tilde{\Ri}}=1/4$ and $\Ri=1/4$ times.

\subsection{The noncritical case}
We now turn to the case in which  the incident wave is not precisely
critical so that $\sigma \neq 0$. We continue with the assumption
that the fluid is inviscid ($\varrho=0$) and that the incident wave
envelope is $A(t_2)=1$ if $t_2>0$.

In this case (\ref{finalequa}) has a steady solution which satisfies the
boundary condition at $\xi=0$:
\begin{equation}
{\cal S}=1 - e^{-\ii \xi/\sigma}.
\label{BillEq:2}
\end{equation}
This steady solution is an approximate version of the  linearly reflected
wave identified by \cite{phillips}. Notice how the scale of the oscillations
is reduced as $\sigma \to 0$. Thus, when the incidence is nearly critical,
one expects to see an initial reduction in scale which is the $t^{-1}$
behaviour identified in (\ref{finalequa}). But this scale reduction is
arrested at time $t_2 ={\cal O}(1/|\sigma|)$ when the Phillips solution in
(\ref{BillEq:2}) is established as a steady state.

In order to understand the details of how the steady solution
in (\ref{BillEq:2})  emerges we can use the solution of the initial
value problem given in (\ref{BillEq:1}):
\begin{subeqnarray}
{\cal S} &=&\int_0^{2\sqrt{\xi t_2}}e^{\ii\sigma u^2\over 4\xi}
 J_1(u)\ \rmn{d}u,\\
{[}{\Psi_0},{W_0}{]}&=&[\Real,\Imag ]\int_0^{2\sqrt{\xi t_2}}
 J_1(u)\ e^{\ii(t+{\sigma u^2\over4\xi}-x)}\  \rmn{d}u,\\
{[}U_0,B_0{]}&=& [-\Real,\Imag ]\int_0^{t_2} J_0
\left(2\sqrt{\xi\tau}\right)e^{\ii(t+\sigma
\tau-x)}\  \rmn{d} \tau,\\
{[}U_1,B_1{]}&=& {1\over2}[-\Real,\Imag ]
\left[e^{2\ii (t-x)}\left({\cal S}{\cal S}_{\xi\xi}-{\cal S}_\xi^2\right)
+\left({\cal S}^\star{\cal S}_\xi\right)_\xi\right],\hskip1truecm\\
 H&=&-\sigma^{-1}\sin\sigma t_2\ e^{-\ii \sigma t_2}.
\end{subeqnarray}

\begin{figure}
\psfig{figure=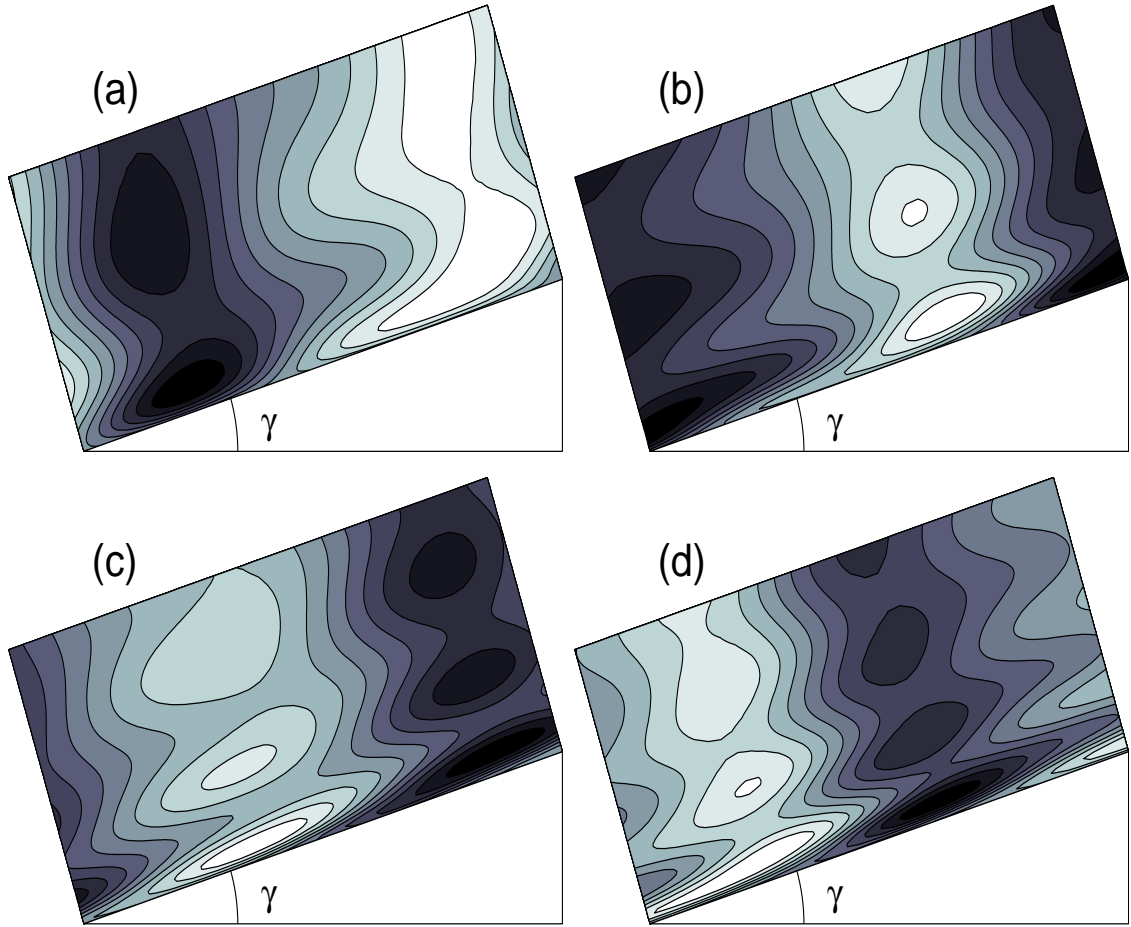,height=15truecm,width=15truecm}
\caption{Streamfunction for four different times
in the  near-critical case: $\beta=22.58^{\circ}$;
$\gamma=17.42^{\circ}$; $\sigma=  -1$; $\varepsilon=0.3$.
$N\sin\beta \ t=5$, 10, 15  and 20 in the panels (a), (b), (c) and (d).
The dimensions of the panel are $\lambda/\sin\alpha$ in the
$x$-direction by  $\lambda/2\sin\alpha$ in the
$z$-direction.}
\label{noncritpsi}
\end{figure}

\begin{figure}
\psfig{figure=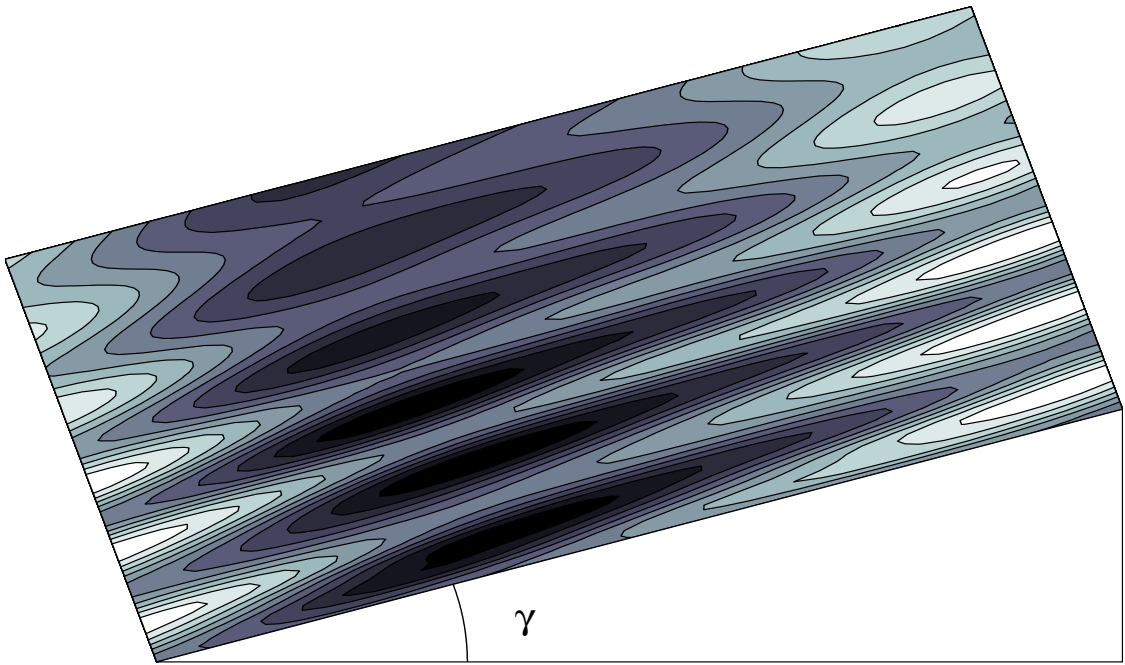,height=15truecm,width=15truecm}
\caption{Streamfunction 
in the  near-critical case at $N\sin\beta \ t=100$: $\beta=22.58^{\circ}$;
$\gamma=17.42^{\circ}$; $\sigma=  -1$; $\varepsilon=0.3$.
The dimensions of the panel are $\lambda/\sin\alpha$ in the
$x$-direction by  $\lambda/2\sin\alpha$ in the
$z$-direction.}
\label{t100}
\end{figure}

\begin{figure}
\psfig{figure=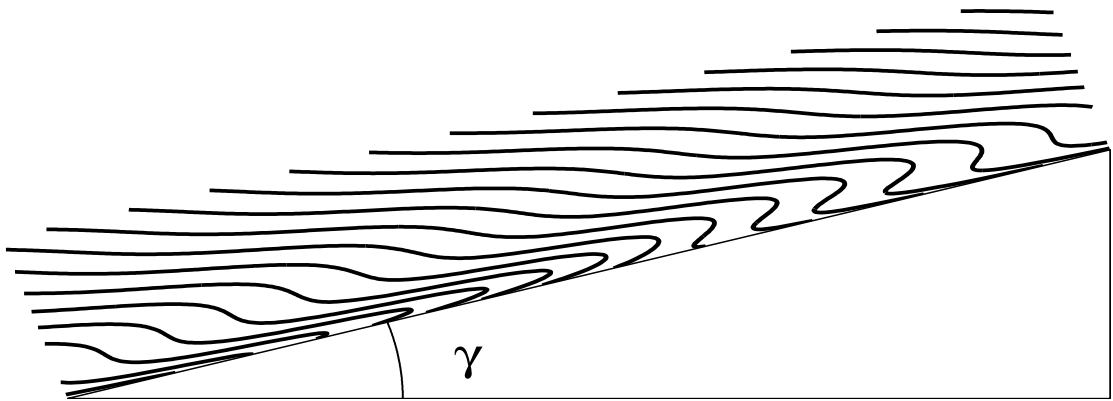,height=15truecm,width=15truecm}
\caption{Buoyancy field  at $N\sin\beta \ t=$ 20
in the near critical case $\beta=22.58^{\circ}$; $\gamma=17.42^{\circ}$ and
 $\varepsilon=0.3$.
The dimensions of the panel are $\lambda/\sin\alpha$ in the
$x$-direction by  $\lambda/5\sin\alpha$ in the
$z$-direction.}
\label{tiltedbuo}
\end{figure}

At early times, the streamfunction in figure~\ref{noncritpsi} is
similar to the critical case in figure~\ref{psi4}. However, as time
progresses, the steady solution in (\ref{BillEq:2}) is set up first in
the neighborhood of the wall, and then this cellular pattern expands
outwards. Figure~\ref{t100} shows the region in which the flow becomes
steady is characterized by regular pattern of vortices.
Figure~\ref{tiltedbuo} shows the isopycnals at a given time: the
pattern is similar to experimental results reported
in figure 12 of~\cite{thorpe} (see also Mac Phee 1998).

\begin{figure}
\null\hskip
3truecm\psfig{figure=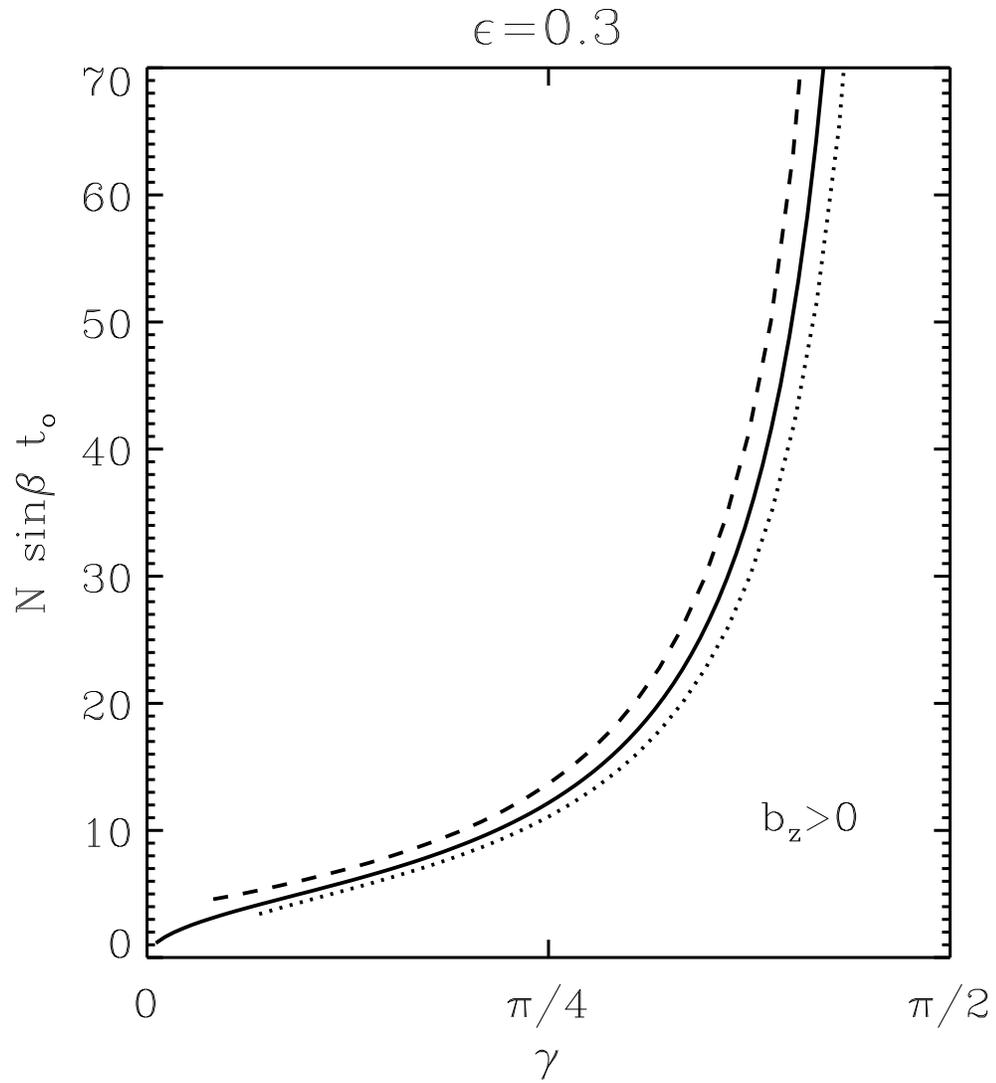,height=15truecm,width=15truecm}
\caption{$N\sin\beta \ t_o$ against the slope
angle $\gamma$. $\sigma=0$ (solid curve), $\sigma=1$ (dotted curve),
$\sigma=-1$ (dashed curve).}  \label{overturnshearsig} \end{figure}

As in the critical case, one can compute $t_o$ and $t_s$
for different values of~$\sigma$. Again, the instability is
convectively driven and initiated at the wall.  $t_o$ is plotted in
 figure~\ref{overturnshearsig} as a function of the slope angle $\gamma$
for different values of $\sigma$.  An interesting point is that, for
a given value of $\gamma$, the positive values of $\sigma$ leads to an
earlier appearance of the instability. This point is
rationalized by noticing that, once $\gamma$ is given, a
positive $\sigma$ corresponds to a smaller value for $\alpha$.  As the
expression (\ref{defto}) is an increasing value of the angle
$\alpha=2\gamma$, near-critical up-slope (respectively down-slope)
reflections are unstable slightly before (respectively after) critical
reflections. Indeed, plotted as a function of $\alpha$, $t_o$ and
$t_s$ are almost independent of $\sigma$.

This is consistent with the observations reported by \cite{imberger97}
that for moderately supercritical waves the instabilities developed
near the bed. They have, however, also studied experimentally the
variation of the  boundary layer thickness as the
incident waves become far from critical, and their results show that
the instability is initiated away from the bed.  In the framework of
this near-critical reflection theory (i.e. for small values of
$\varepsilon$ and $\sigma$), we found that the wave overturning always
starts on the slope, however as time continues the unstable region
extends  away from the boundary (for example, see the unstable region in
 figure~\ref{tiltedbuo}).  For strongly subcritical and supercritical
cases, internal wave reflection from the sloping bed should be
interpreted as wave-wave interaction between the incident and the
reflected waves since, as shown by \cite{imberger97}, the area of
interaction region increases progressively as the waves depart from
critical condition.

\section{Viscous effects}
\label{viscous}

\begin{figure}
\null\vskip -5truecm
\psfig{figure=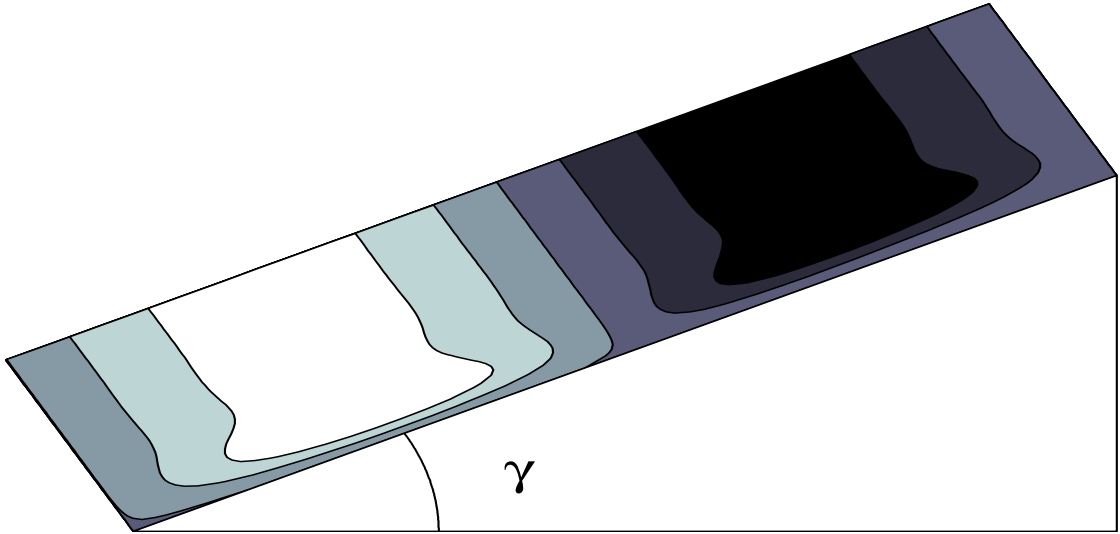,height=15truecm,width=15truecm}
\caption{Final steady state solution
in the critical case: $\beta=32.58^{\circ}$;
$\gamma=27.42^{\circ}$; $\sigma=  -1$; $\varepsilon=0.3$.
The dimensions of the panel are $\lambda/\sin\alpha$ in the
$x$-direction by  $\lambda/4\sin\alpha$ in the
$z$-direction.}
\label{viscoussteady}
\end{figure}

In the viscous case, $\varrho\ne0$,  there is a
steady solution of (\ref{finalequa}) even if the forcing is precisely
critical. For the sake of simplicity,  consider  the case
in which $\varrho \neq 0$, $\sigma=0$, and  the incident wave
envelope is $A(t_2)=1$ if $t_2>0$ (this is a typical experimental switch
on). The steady solution of (\ref{finalequa}) which satisfies the
no-mass flux and no-slip boundary conditions at $\xi =0$ is 
analogous to the western meridional boundary layer, also 
called the Munk layer (see for example Pedlosky 1987). It reads 
\begin{equation}
{\cal S}=1-{2\over \sqrt{3}}
\sin\left({\sqrt3\xi\over 2\varrho^{2/3}}+{\pi\over3} \right)
\ \displaystyle e^{-{\displaystyle \xi/\displaystyle 2\varrho^{2/3}}}.
\label{BillEqn3}
\end{equation}
(we continue to assume that $\kappa=0$ and discard the no
buoyancy flux
boundary condition at $\xi=0$). The solution above,
presented in figure~\ref{viscoussteady} in the extremely viscous
case $\ell=1$,  shows that 
$2\varrho^{2/3}$  is the viscous boundary layer thickness in
the terms of $\xi$; in dimensional variables this  boundary layer
thickness is
\begin{equation}
\ell=\left(\nu+\kappa\over KN\sin\beta\right)^{1/3}
\left(4\over1+\cos\alpha\right)^{1/3}.
\end{equation}
Thus, in the steady state, the viscous boundary layer thickness 
is a decreasing function of the slope angle.

\begin{figure}
\null\hskip 3truecm\psfig{figure=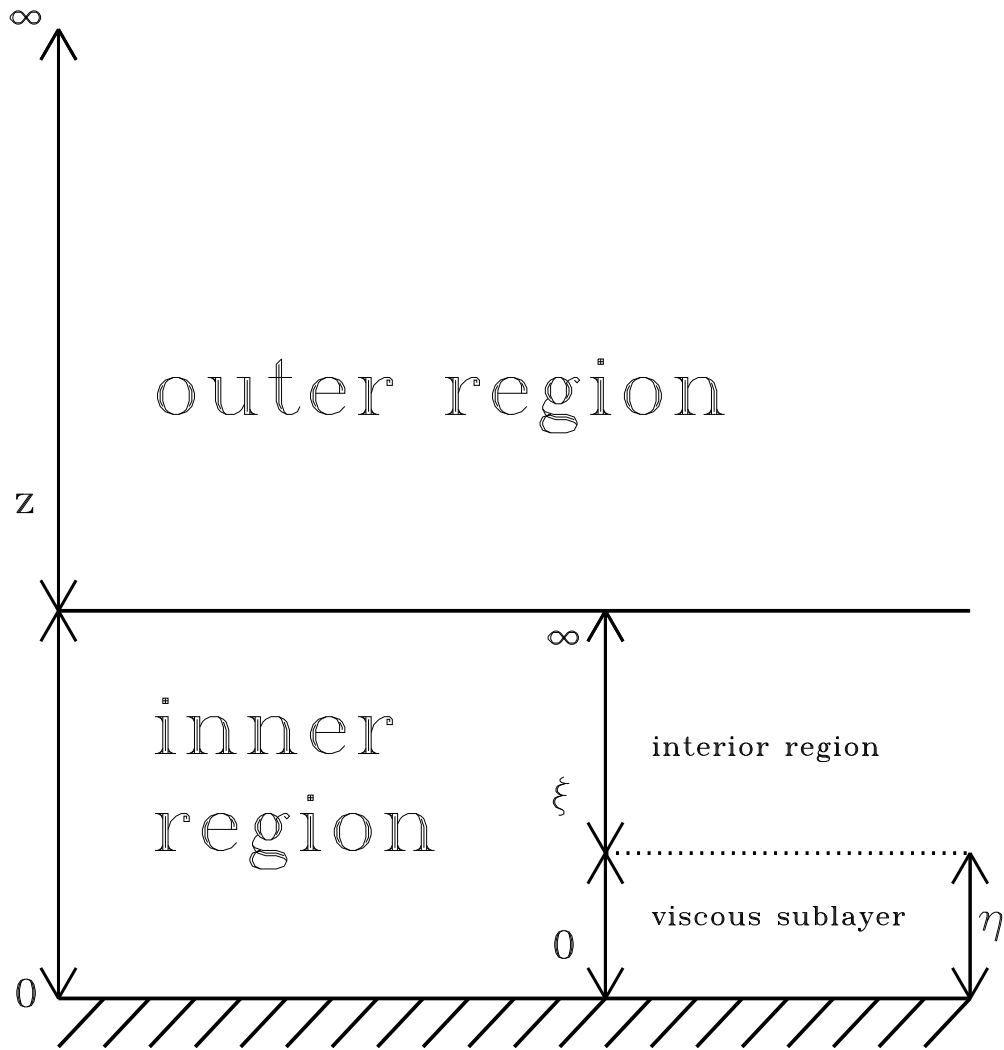,height=15truecm,width=15truecm}
\caption{Schematic view of the different regions used in the
matched asymptotic expansion.}
\label{kundu}
\end{figure}

The steady state solution  is probably irrelevant in many experimental
systems because $\varrho \ll1$ (see table~\ref{table1}) and one expects
that turbulent transition occurs before the steady state is approached.
In this case with $\varrho \ll 1$, we can again use asymptotic matching
to develop an approximate solution of the initial value problem. There
is an interior region in which the effect of viscosity are small and the
solution is approximated  by (\ref{BillEq:1}). However this interior
solution of section \ref{inviscid} does not satisfy the no-slip condition and
so it is necessary to include a  viscous sublayer close to the topography
(see figure~\ref{kundu} for a schematic representation of the different
regions).

The details of this matching problem are in
the appendix. The solution in the interior region is
\begin{eqnarray}
{\cal S}(\xi,t_2)&=& \int_{0}^{t_2} A^*(t_2 -\tau)\ e^{\ii \sigma \tau}
 \sqrt{\xi\over\tau}\
 J_1\left(2\sqrt{\xi \tau}\right)
\ \rmn{d} \tau \nonumber \\
&&+  \varrho 
 \int_0^{t_2} \left({\cal S}_{1\star}^{'}(t_2-\tau)-\ii \sigma{\cal
 S}_{1\star}(t_2-\tau)\right)
 J_0\left(2\sqrt{\xi \tau}\right)
\ \rmn{d} \tau +{\cal O}(\varrho^2)
\label{solutioninterior}
\end{eqnarray}
where
\begin{equation}
{\cal S}_{1\star}(t_2)=-{2\over \sqrt\pi}\int_0^{t_2}\sqrt{t_2-\tau}\
e^{\ii\sigma (t_2-\tau)}\  A^*(\tau) \rmn{d}\tau
\label{definitionoffStar}
\end{equation}
and $J_0$ is the Bessel function of the first kind of order 0.

In the viscous sublayer the solution is
\begin{eqnarray}
{\cal S}(\xi,t_2)&=& \xi \int_0^{t_2} e^{\ii\sigma(t_2-\tau)}
 A^*(\tau)\  \rmn{d} \tau\nonumber\\
&&+\varrho\left[{\cal S}_{1 \star}(\xi,t_2) + \int_0^{t_2} {\rmn d} u
{e^{-\xi^2/4 \varrho^2 u}\over
\sqrt{\pi u}} \int_0^{t_2-u} A^*(\tau)e^{\ii \sigma
(t_2 - \tau)} {\rmn d} \tau
\right] +{\cal O}(\varrho^2).\hskip.5truecm
\label{solutionviscoussublayerforS}
\end{eqnarray}

\begin{figure}
\null\hskip
3truecm\psfig{figure=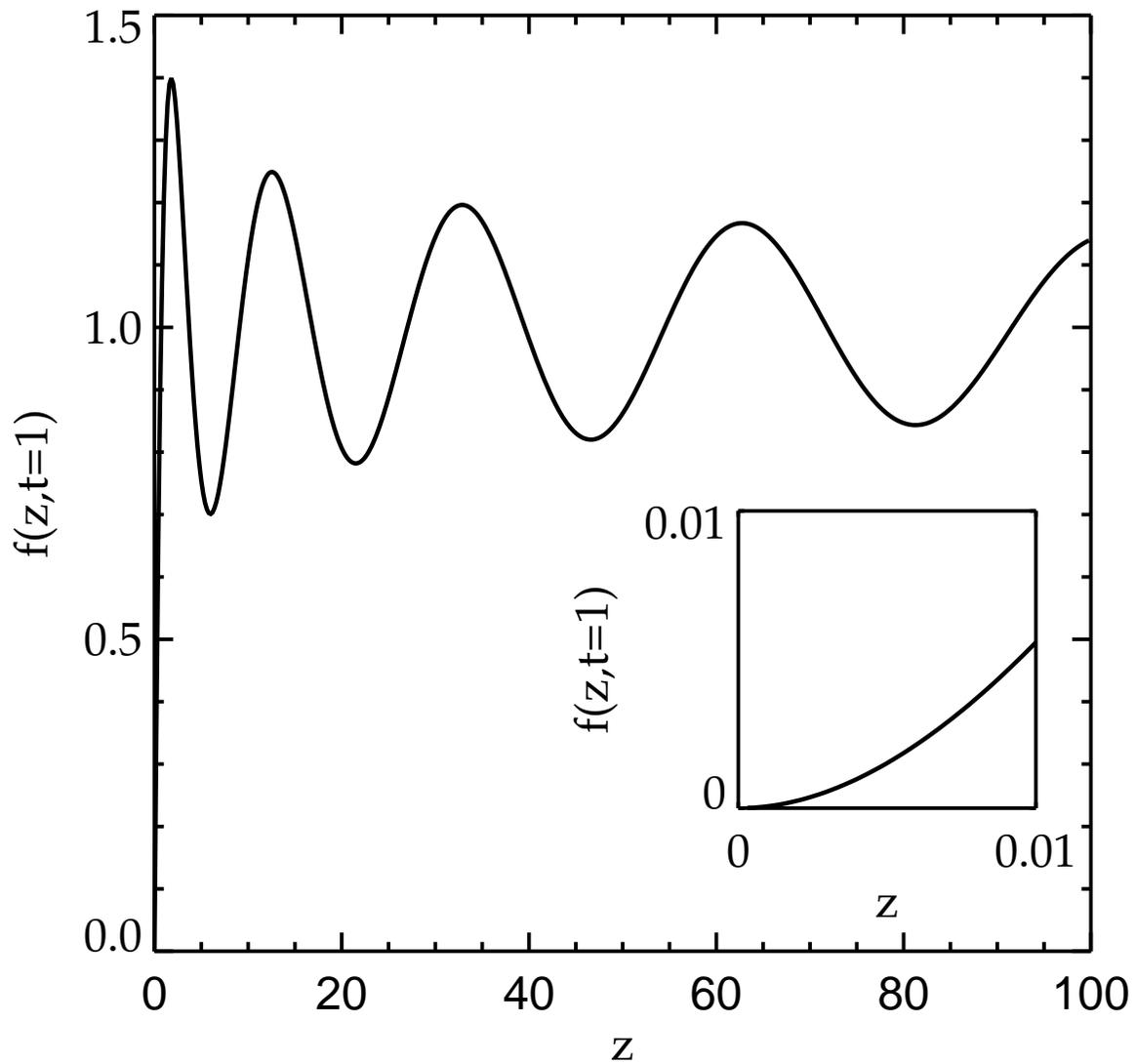,height=15truecm,width=15
truecm}
\caption{Function $f$ against 
the distance normal to the slope $z$
at time $t=1$ and for $\varrho=0.1$, in the critical case.
 In the inset, we plot the viscous sublayer
solution valid only very close to the wall, i.e. $z\ll1$.
}
\label{f_inviscid+vis}
\end{figure}

In the critical case with $\sigma=0$ and $A(t_2)=1$ if $t_2>1$ the
integrals above can be simplified. The interior solution is
\begin{equation}
{\cal S}(\xi,t_2)=1-J_0(2\sqrt{\xi
t_2})-{\varrho\over2\sqrt\pi\xi^{3\over2}}\left(
\sin2\sqrt{\xi t_2}-2\sqrt{\xi t_2} \cos 2\sqrt{\xi
t_2}\right)+{\cal O}(\varrho^2)
\end{equation}
and in the viscous sublayer, the solution   is
\begin{equation}
{\cal S}(\xi,t_2)=\xi t_2-{\varrho t_2^{3\over
2}\over \sqrt\pi}\left[{4\over3}+\int_0^1\rmn{d} u
{u-1\over \sqrt u}\ e^{-{\xi^2/
4\varrho^2t_2u}}\right]+{\cal O}(\varrho^2).
\end{equation}
Both functions are plotted against $z$
in the figure \ref{f_inviscid+vis} in the typical case
$\varrho=0.1$ and $\varepsilon=0.3$. It is clear that the
region where the viscous effects are important corresponds
only to  a very thin region
along the slope.

\section{Conclusion and discussion}
\label{conclusion}

The thrust of this paper has been to study the weakly nonlinear and
nearly critical incidence of internal waves onto a slope. The scalings
of sections 2 and 3 amount to taking the distinguished limit $|\beta
-\gamma| \to 0$ with $a \propto |\beta -\gamma|^{3/2}$ ($a$ is the
amplitude parameter in $(2.12)$). At leading order, these assumptions
give the linear oscillator equation $(3.12a)$ in which the coordinate
normal to the slope, $\xi$, appears only parametrically. Thus, the
buoyancy oscillations along the slope are uncoupled at leading
order. The weak coupling between oscillations at different $\xi$'s is
uncovered by higher orders in the expansion scheme and is apparent in
the forced dispersive wave equation $(3.20)$. One can then view the
incident internal wave as a nearly resonant forcing of this continuum
of weakly coupled (and weakly damped if $\varrho\neq 0$) oscillators.

The scenario above describes the initial evolutionary
stages of nearly critical incidence. However the limitations of this
approach become apparent when the oscillations become so extreme as to
either overturn the buoyancy field, or strongly violate the Miles-Howard
stability condition (see figure 6(c)). In either case, we expect a rapid
transition to turbulence, dramatically enhanced mixing in the
neighbourhood of the slope, and the production of intrusive layers (e.g.,
Ivey \& Nokes 1989; De Silva  1997; Mac Phee 1998).

Within the present framework the most interesting complication which
can be included is {\sl oblique} incidence.  Experimental data by
\cite{eriksen98} and theoretical work by \cite{thorpe97} have recently
emphasized the importance of alongslope currents in the reflection
process for obliquely incident waves in a uniformly stratified
rotating fluid. We speculate that the weakly nonlinear term will
have interesting consequences, such as mean flow induction,
 if the incidence is oblique.

\begin{acknowledgments}
We thank Angel Alastuey, Neil Balmforth, Paola Cessi, Erika McPhee,
Florence Raynal and S. A.  Thorpe for helpful conversations. We also
thank three anonymous referees for their thoughtful review of this
manuscript. This research was supported by the National Science
Foundation under award OCE96-16017 with additional support for TD by a
NATO fellowship. The laboratoire de Physique de l'Ecole Normale
Sup\'erieure de Lyon is URA-CNRS 1325.  \end{acknowledgments}

\appendix
\section{Derivation of the solution in the viscous case.}
\label{appendixviscous}

Defining\refstepcounter{equation}$$
f(\xi,t_2)=\SS(\xi,t_2)\ e^{-\ii\sigma t_2}\qquad\mbox{and}
\qquad g(t_2)=A^*(t_2)\ e^{-\ii \sigma t_2},
\eqno{(\theequation{\itl{a},\itl{b}})}
$$(\ref{finalequa})
becomes
\begin{equation}
f_{t_2\xi}+f-\varrho^2 \  f_{\xi\xi\xi}
=g \label{initialvalueeqvis}
\end{equation}
with  the 4 conditions
\refstepcounter{equation}$$
f(0, t_2 )= f_\xi(0, t_2 )= f_{\xi\xi}(0, t_2 )=0\quad \mbox{and}\quad
\lim_{\xi\rightarrow\infty}\ f(\xi,t_2)=g(t_2).
\eqno{(\theequation{\itl{a}-\itl{d}})}
$$

\subsection{Interior region}
In the interior  region, the  solution of (\ref{initialvalueeqvis})
is obtained expanding $f$ in powers of $\varrho$:
\begin{equation}
f(\xi,t_2)=f_0(\xi,t_2)+ \varrho\ f_1(\xi,t_2) +{\cal O}(\varrho^2).\label{expansionbis}
\end{equation}
The substitution of (\ref{expansionbis}) into (\ref{initialvalueeqvis}) leads
to the following hierarchy:
\begin{subeqnarray}
\varrho^0:\qquad&f_{0t_2\xi}+f_0&=g,\\
\varrho^1:\qquad&f_{1t_2\xi}+f_1&=0.\label{viszero}
\end{subeqnarray}

At leading order, $\varrho^0$, the solution of (\ref{viszero}a)
should satisfy the  2 conditions:
\refstepcounter{equation}$$
f_0(0,t_2)=0\qquad \mbox{and}\qquad
\lim_{\xi\rightarrow\infty}\ f_0(\xi, t_2)=g(t_2).
\eqno{(\theequation{\itl{a},\itl{b}})}\label{condinviscid}
$$
The solution can be obtained using the Laplace transform and we get
\begin{equation}
 f_0(\xi,t_2)=\int_0^{t_2} g(t_2-\tau)\ \sqrt{\xi\over\tau}\
 J_1\left(2\sqrt{\xi \tau}\right)\ \rmn{d} \tau
\label{f_0formula}
\end{equation}
where $J_1$ is the Bessel function of the first kind of order 1.

At  order $\varrho$, (\ref{viszero}b)
with  the conditions
\refstepcounter{equation}$$
f_1(0,t_2)=f_{1\star}(t_2)\qquad \mbox{and}\qquad
\lim_{\xi\rightarrow\infty}\ f_1(\xi, t_2)=0,
\eqno{(\theequation{\itl{a},\itl{b}})}\label{condinviscidun}
$$
leads to the solution
\begin{equation}
f_1(\xi,t_2)
=\int_0^{t_2}
f_{1\star}^{'}(t_2-\tau)\ 
 J_0\left(2\sqrt{\xi \tau}\right)
\ \rmn{d} \tau.
\end{equation}
The unknown function $f_{1\star}(t_2)$ will be determined
 by matching the viscous sublayer solution.

\subsection{Viscous sublayer}

In the viscous sublayer, using the stretched coordinate
$\eta=\xi/\varrho$ and
$h(\eta, t_2)=f(\xi, t_2)/\varrho$,
 (\ref{initialvalueeqvis})
 becomes
\begin{equation}
h_{t_2\eta}-  h_{\eta\eta\eta}+\varrho\ h
=g. \label{initialvalueeqvisin}
\end{equation}
Expanding $h$ in powers of $\varrho$,
the leading order solution is
\begin{equation}
h_{0}(\eta,t_2)=f_{1\star}(t_2)+\eta\int_0^{t_2} g(\tau)\rmn{d}  \tau
+r(\eta,t_2),
\end{equation}
where $r(\eta,t_2)$ satisfies the diffusion equation
\begin{equation}
r_{t_2}=r_{\eta\eta},\label{heatequation}
\end{equation}
with the three conditions
\refstepcounter{equation}$$
r_\eta(0,t_2)=-\int_0^{t_2}g(\tau)\rmn{d}  \tau,
\qquad  \lim_{\xi\rightarrow\infty}r=0\quad \mbox{and}\quad
r(0,t_2)=-f_{1\star}(t_2).
\eqno{(\theequation{\itl{a}-\itl{c}})}
$$
Let us stress that the two first conditions define
completely the solution of (\ref{heatequation}),
whereas the last one determines $f_{1\star}(t_2)$ in terms of $g(t_2)$.

Using the Laplace transform, one obtains the solution of
(\ref{heatequation}) and we finally get, not only the  function
\begin{equation}
f_{1\star}(t_2)=-{2\over \sqrt\pi}\int_0^{t_2}\sqrt{t_2-\tau}\
 g(\tau) \rmn{d}\tau
\label{definitionoffstar}
\end{equation}
 but also the the general
solution in the viscous sublayer given
by 
\begin{equation}
f(\xi,t_2)=\xi\int_0^{t_2} g(\tau)\rmn{d}  \tau+\varrho
\left[f_{1\star}(t_2)+
\int_0^{t_2}\rmn{d} u\
\displaystyle {e^{-{\xi^2/4 \varrho^2u}}\over
\sqrt{\pi u}}\ \int_0^{t_2-u} g(\tau)\rmn{d}\tau\right] +{\cal O}(\varrho^2).
\label{solutionviscoussublayer}
\end{equation}

\vfill\eject

\vfill\eject

\vfill\eject

\vfill\eject
\vfill\eject

\vfill\eject

\vfill\eject
\vfill\eject
\vfill\eject

\vfill\eject
\vfill\eject
\vfill\eject
\vfill\eject
\vfill\eject

\end{document}